\documentclass{article}

\usepackage[version=4]{mhchem}
\usepackage{enumitem}
\usepackage{siunitx} 
\usepackage{bm}
\usepackage{caption}
\usepackage{array}
\usepackage{xcolor}

\usepackage{amssymb}
\usepackage{upgreek}

\usepackage{arxiv}

\usepackage[utf8]{inputenc} 
\usepackage[T1]{fontenc}    
\usepackage{hyperref}       
\usepackage{url}            
\usepackage{booktabs}       
\usepackage{amsfonts}       
\usepackage{nicefrac}       
\usepackage{microtype}      
\usepackage{cleveref}       
\usepackage{lipsum}         
\usepackage{graphicx}
\usepackage{natbib}
\usepackage{doi}

\title{Direct numerical simulation of thermo-diffusively unstable premixed hydrogen-air flames in a fully-developed turbulent channel flow at ${Re_\tau=530}$}

\newif\ifuniqueAffiliation
\uniqueAffiliationtrue

\ifuniqueAffiliation 
\author{ \href{https://orcid.org/0000-0002-7837-0939}{\includegraphics[scale=0.06]{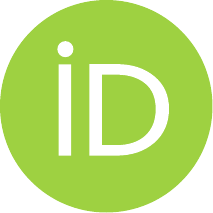}\hspace{1mm}Felix Rong}\thanks{Corresponding author: rong@stfs.tu-darmstadt.de} \\
	Technical University of Darmstadt\\
    Department of Mechanical Engineering\\
    Simulation of Reactive Thermo-Fluid Systems\\
    Otto-Berndt-Str. 2, 64287 Darmstadt, Germany
	\And
	\href{https://orcid.org/0000-0001-5166-6273}{\includegraphics[scale=0.06]{orcid.pdf}\hspace{1mm}Max Schneider} \\
    Technical University of Darmstadt\\
    Department of Mechanical Engineering\\
    Simulation of Reactive Thermo-Fluid Systems\\
    Otto-Berndt-Str. 2, 64287 Darmstadt, Germany
	\And
	\href{https://orcid.org/0000-0002-0355-2252}{\includegraphics[scale=0.06]{orcid.pdf}\hspace{1mm}Hendrik Nicolai} \\
    Technical University of Darmstadt\\
    Department of Mechanical Engineering\\
    Simulation of Reactive Thermo-Fluid Systems\\
    Otto-Berndt-Str. 2, 64287 Darmstadt, Germany
	\And
	\href{https://orcid.org/0000-0001-9333-0911}{\includegraphics[scale=0.06]{orcid.pdf}\hspace{1mm}Christian Hasse} \\
    Technical University of Darmstadt\\
    Department of Mechanical Engineering\\
    Simulation of Reactive Thermo-Fluid Systems\\
    Otto-Berndt-Str. 2, 64287 Darmstadt, Germany
	\And
	\href{https://orcid.org/0000-0003-2753-9690}{\includegraphics[scale=0.06]{orcid.pdf}\hspace{1mm}Andrea Gruber} \\
    SINTEF Energy Research, Thermal Energy Department, 7465 Trondheim, Norway
    \\[2pt]
    Norwegian University of Science and Technology\\
    Department of Energy and Process Engineering, 7491 Trondheim, Norway
    }

\begin{document}
\maketitle

\begin{abstract}
Direct Numerical Simulations (DNS) of premixed hydrogen-air flames anchored in a fully-developed turbulent channel flow (TCF) are performed at a friction Reynolds number of $\mathrm{Re}_\tau=530$ and thermochemical conditions susceptible to the emergence of intrinsic thermo-diffusive (TD) phenomena acting on the turbulent flame. Two premixed flames are studied: a slower flame ($\varphi=0.25$), predominantly propagating within the core flow, and a faster one ($\varphi=0.35$), reaching closer to the channel walls and intermittently quenching on it.

The present DNS database provides new insights into the characteristics of premixed flames susceptible to TD phenomena and propagating in realistic near-wall shear turbulence. The influence of varying turbulence intensity, and of wall-distance dependent time and length scales, on the flame propagation characteristics is evaluated through a detailed analysis of the local stretch factor $I_0$, quantifying reactivity enhancements caused by TD phenomena.

At $\varphi=0.25$, the flame response to the fluid motions is mainly forced by the weaker turbulence present in the core flow. This results in an augmented $I_0$ compared to the laminar reference value, suggesting reactivity enhancement by the strongly non-linear interaction of TD phenomena with (relatively) weak turbulent motions present within the core flow. At $\varphi=0.35$, as the flame propagates from the core flow towards the channel walls, the flame response is forced by turbulence of increasing intensity, resulting in a corresponding augmentation of the Karlovitz number. Crucially, as the flame propagates into the near-wall region, the peak value of $I_0$ is co-located with the peak Reynolds stresses ($y^+ \sim 10$). This observation suggests a strong (local) synergistic interaction between TD phenomena and wall turbulence, ultimately resulting in significantly enhanced flame speed.
\end{abstract}



\section{Introduction} \label{sec: intro}

Hydrogen, as a carbon-free energy carrier, holds significant promise for low-emissions combustion applications \citep{dreizler_role_2021, pitsch_transition_2024}. In particular, premixed hydrogen flames under fuel-lean conditions are of technical relevance because of their reduced flashback propensity and low $\si{NO}_x$ emissions. 
However, such fuel-lean hydrogen flames are subject to thermo-diffusive (TD) phenomena that are intrinsic to the flame front and that can strongly influence flame reactivity and combustion characteristics beyond well-established scaling laws from combustion of natural gas \citep{pitsch_transition_2024}.

The unique molecular and thermophysical properties of hydrogen, particularly its fast molecular diffusion relative to other species (preferential diffusion) and heat (differential diffusion), quantified by its low Lewis number ($Le_\mathrm{H_2} \approx 0.3$), cause cellular instabilities in laminar flames that were reported and investigated already decades ago \citep{zeldovich_1944,markstein_1949}. A detailed explanation of the physical processes leading to the occurrence of these cellular instabilities is provided by \citet{williams_combustion_1985} that refers to them as \textit{diffusive-thermal (DT) instabilities}. More recent literature frequently utilize the alternative (and equivalent) nomenclature \textit{thermo-diffusive (TD) instabilities}, arising in \textit{TD-susceptible flames} and caused by \textit{TD phenomena}. The latter nomenclature is adopted in the present manuscript. Because of preferential and differential diffusion, molecular hydrogen is locally transported into positively curved regions of the flame (convex towards the reactants) faster than other reactants and also faster than heat is transported out of the same regions. Conversely, hydrogen concentration is locally reduced in negatively curved regions (convex towards the products). This imbalance in mass and heat diffusion represents the main driver of TD instabilities: the flame propagates at a locally varying mixture fraction and flame speed, with richer regions burning hotter and faster, thus destabilizing the global flame front and creating cellular structures that penetrate into the unburnt gas \citep{berger_characteristic_2019, howarth_empirical_2022}. These TD phenomena and instabilities have been extensively investigated through theoretical, experimental, and numerical studies in laminar flow configurations \citep{MATALON2003, Altantzis2012, berger_characteristic_2019, howarth_empirical_2022, Lulic2023, lapenna_synergistic_2024}.

In turbulent flows, the additional effect of turbulent motions on flame-front wrinkling is superimposed, in a strongly non-linear interaction, onto the intrinsic TD phenomena, simultaneously acting on it, and potentially leading to a significant increase in the global burning rate. This effect has been reported by several research groups and is also referred to as 'synergistic interaction' in some studies \citep{aspden_turbulenceflame_2011,aspden_towards_2019,berger_synergistic_2022, berger_effects_2024, yao_isolating_2024}.
In such cases, the enhancement in turbulent flame speed and global burning rate cannot be solely attributed to the increase in flame surface area caused by turbulence, as for more conventional gaseous fuels (e.g. natural gas), nor to TD phenomena alone, but represents the outcome of non-linear interactions between these two processes.
Several Direct Numerical Simulation (DNS) studies of premixed hydrogen-air combustion have considered TD-susceptible flames in statistically planar inflow–outflow configurations subjected to spectrally-forced or decaying isotropic turbulence \citep{aspden_turbulenceflame_2011, aspden_turbulence-chemistry_2015, howarth_thermodiffusively-unstable_2023, rieth_effect_2023, yao_isolating_2024, hunt_thermodiffusively-unstable_2025}. While these isotropic-turbulence configurations provide detailed insights about the response of TD-susceptible planar flames to very specific turbulence characteristics, they are not able to correctly capture the flame response to the anisotropic structure and variance of characteristic length and time scales occurring in realistic shear-generated turbulence.

In practical combustion applications, complex flow patterns and confinements typically cause the interaction of flames with turbulent flows that are characterized by strongly anisotropic turbulent velocity fields. Consequently, recent DNS studies have considered TD-susceptible flames in more realistic shear-turbulence configurations.
Examples include the turbulent slot flames studied by \citet{berger_synergistic_2022, berger_effects_2024} and the turbulent shear layers analyzed by \citet{rieth_enhanced_2022}. 
However, to date, no DNS study has investigated the propagation characteristics of TD-susceptible premixed hydrogen flames interacting with the structure of near-wall shear turbulence in fully-developed boundary layers.

\begin{figure}[ht!]
    \centering
    \includegraphics[width=0.525\linewidth]{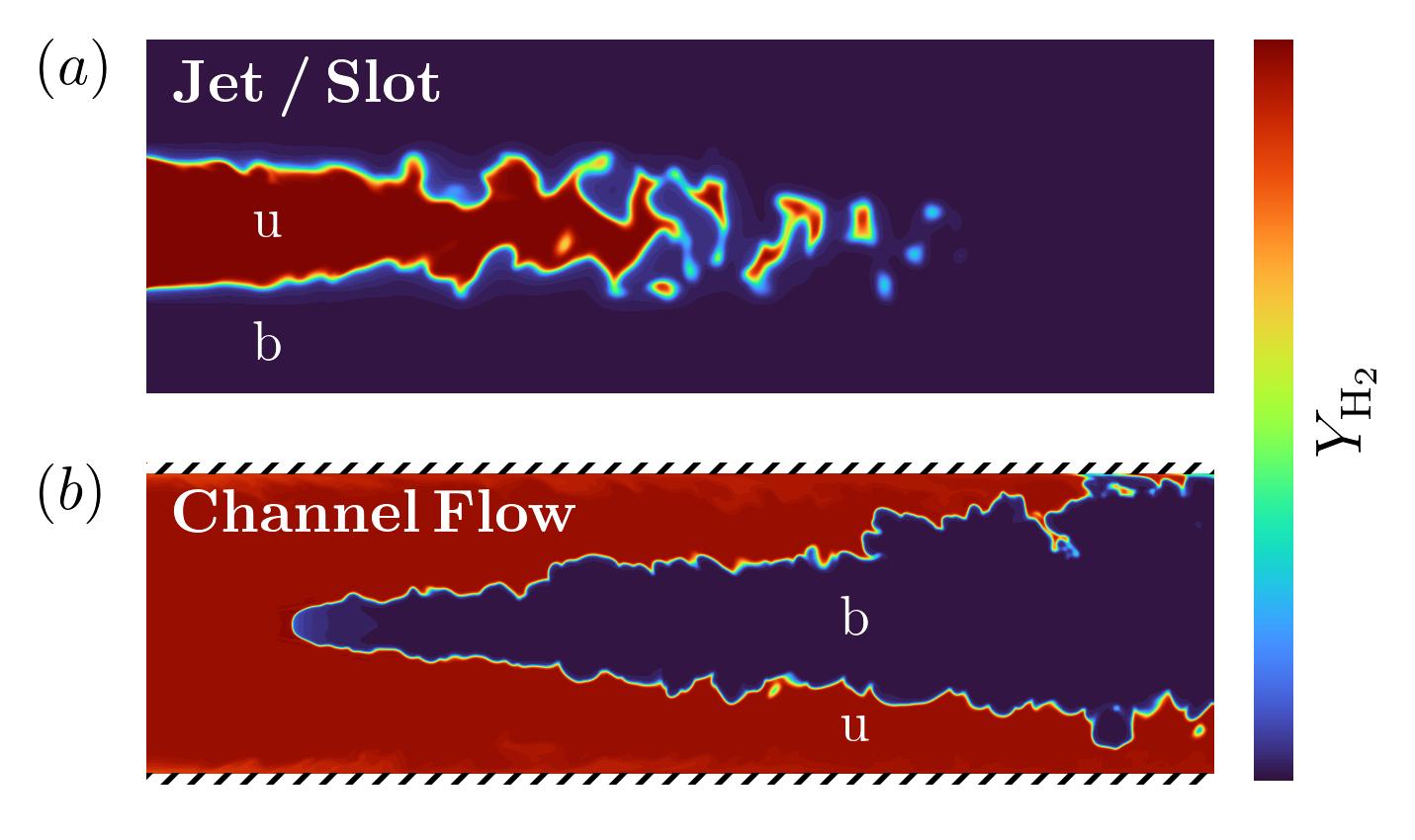}
    \caption{$(a)$ Turbulent jet flame, taken from \citet{Nicolai2025}; $(b)$ V-shaped flame anchored in a TCF configuration, with hatched lines indicate the channel walls. The flames are colored by the hydrogen fuel mass fraction $Y_{\ce{H2}}$, and the unburnt and burnt regions are denoted by 'u' and 'b', respectively.}
    \label{fig: Jet Channel flame}
\end{figure}

The fully-developed turbulent channel flow (TCF) is a canonical wall-bounded flow configuration that is well characterized from a fluid mechanics perspective and enables the investigation of turbulent premixed flame propagation in realistic, shear-driven turbulence.
Figure~\ref{fig: Jet Channel flame}$(a)$ illustrates a turbulent jet/slot flame colored by the fuel mass fraction, while figure~\ref{fig: Jet Channel flame}$(b)$ shows a V-shaped flame anchored in a TCF configuration. These images qualitatively compare the global morphological similarities and the complementary characteristics of two flame configurations featuring spatially reversed unburnt and burnt regions. Evidently, the DNS of the V-shaped premixed hydrogen flames examined in the present TCF setup serves as a valuable complement to earlier DNS studies of jet and slot flames conducted at comparable Reynolds numbers \citep{berger_effects_2024}.

The TCF configuration with an anchored V-shaped flame, used here to examine the propagation characteristics of thermo-diffusively susceptible flames interacting with realistic shear-driven turbulence under statistically steady conditions, has been employed extensively in previous studies, most notably for investigations of turbulent flame--wall interaction (FWI), a topic of considerable practical relevance in combustion systems \citep{Dreizler2015}. Although turbulent FWI remains an active research area with significant implications for flame stability, combustion efficiency, and pollutant formation \citep{Dreizler2015}, it is not the focus of the present study.

One of the earliest DNS investigations of a V-shaped flame anchored in turbulent Couette flow at relatively low Reynolds numbers was conducted by \citet{alshaalan_vflame_2002}, who employed a single-step chemical model to study FWI. A few years later, \citet{gruber_turbulent_2010} performed a DNS of a premixed hydrogen–air flame anchored in turbulent Poiseuille flow at a friction Reynolds number of $Re_\tau = 180$ under fuel-rich conditions ($\varphi = 1.5$), using detailed chemical kinetics and transport. As is common in experimental and numerical TCF studies, the friction Reynolds number $Re_\tau = h / \delta_\nu$, defined by normalizing the channel half-width with the viscous length scale, is used throughout the following discussion to characterize the confined turbulent flow (see next section for details).

These initial FWI studies, employing anchored V-shaped flames, revealed key differences between single-step and detailed chemistry models, and highlighted the role of streamwise-elongated vortical structures in the turbulent boundary layer in governing flame propagation and quenching. Subsequent DNS efforts shifted attention away from FWI-induced quenching and toward the dynamics of near-wall flame propagation during boundary-layer flashback in hydrogen–air flames \citep{gruber_direct_2012,gruber_direct_2015,chen_BLF_2023}, including the influence of inlet mixture stratification \citep{gruber_direct_2018}.

Computational constraints limited most early DNS studies to comparatively low friction Reynolds numbers ($Re_\tau \sim 180$), restricting the scale separation and extent of the inertial subrange that could be achieved. More recent DNS work has extended TCF investigations to higher friction Reynolds numbers ($Re_\tau = 313$–$395$), enabling the analysis of propagation and quenching characteristics of ammonia- and methane-air flames \citep{chi_effect_2024,kai_analysis_2025}.

However, among these earlier efforts, only few studies have focused on hydrogen flames and, crucially, even fewer on thermochemical conditions susceptible to the occurrence of TD phenomena. For instance, while featuring realistic three-dimensional turbulence and detailed chemical kinetics, the FWI investigation of \citet{gruber_turbulent_2010} did not consider TD-susceptible flames. More recently, \citet{PartI, PartII} examined laminar FWI of TD-susceptible hydrogen flames but this study was limited to laminar head-on quenching (HOQ) in a two-dimensional (2D) configuration.

To date, no full-fledged DNS study has examined in detail the propagation characteristics of premixed hydrogen flames susceptible to thermo-diffusive (TD) phenomena while nonlinearly interacting with realistic near-wall turbulence at elevated Reynolds numbers.

The present work addresses this gap by simulating fuel-lean premixed hydrogen–air flames anchored in a fully developed turbulent channel flow at $Re_\tau = 530$ and at thermochemical conditions prone to TD effects ($\varphi = 0.25$ and $\varphi = 0.35$), using detailed chemical kinetics and transport. This friction Reynolds number substantially exceeds those used in previous DNS studies of turbulent reactive channel flows, enabling a wider range of dynamically relevant scales, including a more extensive inertial subrange, and providing a high-fidelity reference database for the development and validation of turbulent combustion models applicable to shear flows near solid boundaries. In addition, this investigation provides the first detailed assessment of TD-susceptible hydrogen flame propagation in a canonical wall-bounded shear flow characterized by strong anisotropy and realistic near-wall turbulence at such (relatively) high $Re_\tau$. The pronounced variation in turbulence intensity with wall distance leads to localized changes in the Karlovitz number, which modulate the nonlinear interplay between turbulence and TD effects and, in turn, shape the global flame-propagation behavior.

Although the primary focus is on flame-propagation dynamics in the presence of near-wall turbulence, one of the simulated conditions ($\varphi = 0.35$) also exhibits intermittent flame quenching at the solid surface. This behavior provides an opportunity to complement insights from earlier two-dimensional studies \citep{PartI, PartII}, extending them to a fully three-dimensional turbulent environment.

In summary, the present study aims to answer the following research questions:
\begin{enumerate}[%
    label=(\roman*),      
    style=multiline,          
    leftmargin=3em,        
    itemsep=0.15\baselineskip  
]
  \item What are the global flame propagation characteristics of TD-susceptible premixed hydrogen-air flames interacting with realistic near-wall turbulence?
  \item  How is the local topology and microstructure of the TD-susceptible flames affected by the turbulent motions as a function of the wall-normal distance?
  \item  How does the non-linear interaction between near-wall turbulence and TD phenomena vary the stretch factor $I_0$ with the wall-normal distance and with the associated variation in Karlovitz number?
\end{enumerate}
\vspace{0.25em}

The remainder of this paper is structured as follows. Section~\ref{sec: setup} describes the numerical setup and computational methodology. Section~\ref{sec: results} presents the DNS results and analysis. The study concludes with a summary of the main findings in section~\ref{sec: conclusion}.


\section{Computational setup} \label{sec: setup}

\begin{figure}[ht!]
    \centering
    \includegraphics[width=0.625\linewidth]{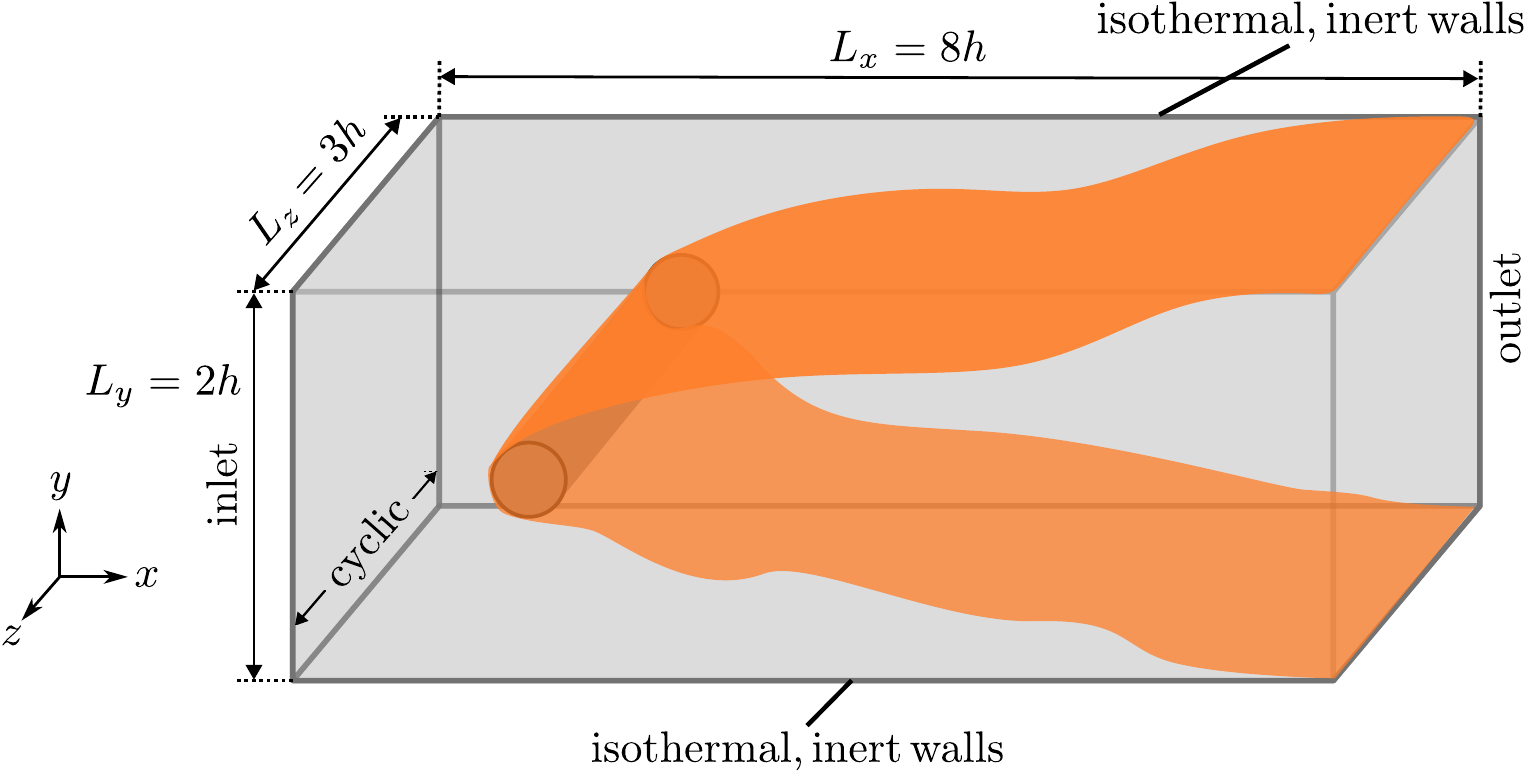}
    \caption{
    Schematic representation of the simulated V-shaped flame anchored in a TCF configuration. The orange surface denotes the flame front, and the flame anchor is depicted as a cylinder.
    }
    \label{fig: numerical setup}
\end{figure}

The configuration used in this study builds on the numerical setup of \citet{gruber_turbulent_2010} to investigate the structure and propagation of fuel-lean premixed hydrogen-air flames susceptible to TD phenomena, extending the configuration to a higher friction Reynolds number of $\mathrm{Re}_\tau = h / \delta_\nu = 530$ (up from $180$) where $h$ is the channel half-width and $\delta_\nu$ is the viscous length scale \citep{Pope2000}.
Figure~\ref{fig: numerical setup} shows a schematic of the computational setup, including boundary conditions. Three-dimensional (3D) rendered videos of the flames are provided Movie~1 and Movie~2 in the supplementary material.

A premixed fuel/oxidizer mixture enters the computational domain at its inlet boundary ($x=0$) and the flame stabilizes at the flame holder (shown as a cylinder in figure~\ref{fig: numerical setup}), propagating downstream and outward toward the walls in the form of a V-shaped surface.
The streamwise, wall-normal and spanwise directions are denoted as $x$, $y$ and $z$, respectively, and non-reflective Navier-Stokes characteristic boundary conditions are applied \citep{Poinsot1992, Sutherland2003} at the inflow ($x=0$) and outflow ($x=L_x$) planes.
The no-slip walls at $y=0$ and $y=L_y$ are assumed to be perfectly smooth, isothermal and inert surfaces, with zero wall-normal fluxes of the chemical species.
Periodic boundary conditions are applied in the homogeneous spanwise direction ($z=0$ and $z=L_z$).
A flame holder is required because the mean flow velocity is considerably larger than the turbulent flame speed \citep{gruber_turbulent_2010}, and it is modeled through a fixed energy deposit. 
This approach neither influences the flow field nor the local flame characteristics sufficiently far from the flame holder and outside of the hot core.
The flame holder has a diameter of $1\,\ce{mm}$ and this approximately corresponds to $53$ wall units, i.e., lengths normalized by the viscous scale $\delta_\nu$ as $x^+ = x / \delta_\nu$,  $y^+ = y / \delta_\nu$ and $z^+ = z / \delta_\nu$. Its center is located at $x_{\text{anc}}^+ = y_{\text{anc}}^+ = 530$ on the channel centerline, equidistant from the walls and the inlet and spanning the homogeneous direction.
Additional details on the DNS initialization and the numerical solver are provided in section~\ref{sec: solver}.

The flow parameters, channel geometry, and numerical grid resolution of the DNS are presented in table~\ref{tab: computational setup}. The computational domain is chosen to be considerably larger than the minimal channel requirements \citep{Jimnez1991}.
A rectilinear computational mesh is employed which is uniform in both the streamwise and spanwise directions. The mesh is mildly stretched in the wall-normal direction to resolve the near-wall boundary layer, but remains sufficiently fine towards the core flow to resolve the thin flame front. There are two grid points within $y^+=1$ and 16 grid points within $y^+=10$, satisfying the recommended resolution for viscous sublayers \citep{moser_direct_1999}. 
The nominal friction Reynolds number of $Re_\tau = 530$ corresponds to a bulk Reynolds number of $Re_\mathrm{m} \approx 9{,}625$ and a centerline Reynolds number of $Re_\mathrm{c} \approx 11{,}000$ for the approaching flow. 

\begin{table}[ht!]
\centering
\small 
\setlength{\tabcolsep}{7pt} 
\renewcommand{\arraystretch}{1.3} 
\addtolength{\aboverulesep}{2pt}
\addtolength{\belowrulesep}{2pt}
\begin{tabular}{@{}p{0.25\linewidth} p{0.45\linewidth} p{0.2\linewidth}@{}}
    \toprule
    $Re_\tau = 530$ & $L_x \times L_y \times L_z = 8h \times 2h \times 3h$ & $h=1\,\mathrm{cm}$ \\ 
    $T_\mathrm{w} = T_\mathrm{u} = 750 \,\mathrm{K}$ & $N_x,\,N_y,\,N_z = 3200,\,750,\,1200$ & $U_\tau=2.19\,\mathrm{m/s}$ \\
    $p=2\,\mathrm{bar}$ 
    & $\Delta x^+,\,\Delta y^+,\,\Delta z^+ = 1.33, \, 0.62\mathrm{-}1.51, \,1.33$ & $\delta_v = 18.86\,\mu\mathrm{m}$ \\
    \bottomrule
\end{tabular}
\vspace{3mm}
\caption{Parameters of the DNS setup, where $h$ is the channel half-width.}
\label{tab: computational setup}
\end{table}

The detailed chemical kinetics scheme for hydrogen oxidation developed by \citet{Li2004} is employed, which consists of 9 species and 19 elementary reactions. Nitrogen is treated as an inert species and chemical pathways leading to formation of nitrogen oxides are omitted. No heterogeneous or catalytic surface reactions are included, assuming chemically-inert walls. Although the assumption of non-reactive solid surfaces, and the resulting overprediction of near-wall radical concentrations, has been shown to overestimate heat-release rates during FWI and quenching of stoichiometric or fuel-rich flames \citep{de_nardi_infinitely_2024}, this assumption does not adversely affect the fuel-lean conditions considered here. A mixture-averaged diffusion model is used (CHEMKIN and TRANSPORT libraries \citep{Kee1989}) to capture the effects of differential and preferential diffusion, which are particularly important for hydrogen flames \citep{pitsch_transition_2024}. 
The Ludwig-Soret effect (thermodiffusion) is also included, as it is known to have significant quantitative impact on hydrogen flames, amplifying the effects of TD phenomena \citep{Schlup2017, Howarth2024, Zirwes2024}.

Two premixed flames are examined: the first configuration consists of an ultra-lean, homogeneous mixture of air and hydrogen burning at an equivalence ratio of $\varphi=0.25$, primarily propagating in the core flow; the second configuration has a nominal equivalence ratio of $\varphi = 0.35$, with a flame that propagates faster and extends closer to the walls, featuring near-wall combustion and intermittent quenching/FWI. In the latter configuration, to represent the near-wall compositional variance often found in gas turbines burners, the mixture is mildly stratified at the inlet plane with a leaner layer entering the domain in the vicinity of the wall. In the following, the two cases are referred to by their nominal equivalence ratios ($\varphi=0.25$ and $\varphi=0.35$).
The fresh gas temperature is chosen to match the nominal temperature in large stationary gas turbines for power generation. The pressure is limited to $p=2\,\mathrm{bar}$ due to the extreme computational costs required to fully resolve the flame structure at the high-pressure conditions present in gas turbines (typically exceeding \SI{10}{bar}).
Table~\ref{tab: 1D flames} summarizes the nominal properties of homologous one-dimensional (1D) laminar, freely propagating (FP) flames. Both turbulent flames considered here are expected to be susceptible to TD phenomena under the present conditions, as their effective Lewis numbers fall below the critical threshold given by linear stability theory \citep{matalon_intrinsic_2007} and the parameter $\omega_2$ is positive \citep{howarth_empirical_2022}. The latter quantity is based on the theoretical work of \citet{MATALON2003} and can quantify the TD response as presented by \citet{howarth_empirical_2022, howarth_thermodiffusively-unstable_2023}.
In addition, linear stability analysis conducted numerically confirms that the thermochemical conditions of both flames are susceptible to TD phenomena, as shown in the supplementary document section S1.1. Table~\ref{tab: 1D flames} also provides estimates of the nominal \textit{near-wall} Damköhler number ($Da^0=t_\mathrm{w}/\tau_{_\mathrm{F}}$), which characterizes the ratio between the timescales of flow and chemical reaction and it is calculated here from the nominal laminar flame time $\tau_{_\mathrm{F}} = \delta_\mathrm{F} / s_\mathrm{L}$ and the wall time $t_\mathrm{w} = \nu / U_\tau^2$ of the non-reactive TCF configuration \citep{gruber_turbulent_2010,gruber_direct_2018}.

\begin{table}[ht!] \centering \small
\renewcommand{\arraystretch}{1.3} 
\setlength{\tabcolsep}{15pt} 
\centerline{
\addtolength{\aboverulesep}{2pt}
\addtolength{\belowrulesep}{2pt}
\begin{tabular}{lcc}
\toprule
  & $\bm{\varphi=0.25}$ & $\bm{\varphi=0.35}$ \\[1mm]
\midrule
Equivalence ratio $\varphi$ & 0.25 & 0.35 \\
Laminar burning velocity $s_\mathrm{L}\,/\,\si{[m/s]}$ & 1.27 & 3.08 \\
Thermal flame thickness $\delta_\mathrm{F}\,/$ \si{[\mu m]} & 260 & 212 \\
Adiabatic flame temperature $T_\mathrm{b}\,/$ \si{[K]} & 1462 & 1703 \\
Effective Lewis number $Le_\mathrm{eff}$ & 0.38 & 0.43 \\
Critical Lewis number $Le_\mathrm{crit}$ & 0.73 & 0.60 \\
Instability parameter $\omega_2$ & 3.10 & 1.20 \\
Nominal Damköhler number $Da^0$ & 0.042 & 0.129 \\
Computational costs $/$ $10^6$ \si{CPUh} & 81.84 & 50.11\\
\bottomrule 
\end{tabular}
}
\vspace{3mm}
\caption{
Characteristic flame properties of the homologous 1D laminar FP flames representative of configurations $\varphi=0.25$ and $\varphi=0.35$. The thermal flame thickness is defined as $\delta_\mathrm{F}=(T_\mathrm{b}-T_\mathrm{u})\big/\max(\mathrm{d}T/\mathrm{d}x)$, where the subscripts 'b' and 'u' denote the burnt and unburnt mixture, respectively.
The definitions of the effective Lewis number $Le_\mathrm{eff}$, critical Lewis number $Le_\mathrm{crit}$ and instability parameter $\omega_2$ can be found in the work of \citep{howarth_empirical_2022}.
}
\label{tab: 1D flames}
\end{table}


\subsection{DNS Solver} \label{sec: solver}
The simulations are performed using the DNS solver S3D \citep{Chen2009}. The algorithm solves the Navier–Stokes equations for a compressible reactive fluid on a structured Cartesian mesh using high-order numerical schemes. Spatial derivatives are computed using an eighth-order explicit centered finite difference scheme (with third-order one-sided stencils used at the domain boundaries), and a tenth-order explicit spatial filter is employed to remove high-frequency noise and reduce aliasing error. Time integration is performed with a fourth-order six-stages explicit Runge–Kutta method.
S3D is implemented in FORTRAN 90 and parallelized using the Message Passing Interface (MPI) for inter-process communication in parallel execution. The simulations, parallelized using a domain decomposition of 80 × 25 × 30 MPI ranks, were carried out on the Betzy HPC facility, located at NTNU in Trondheim, using 60~000 AMD CPU cores (500 nodes). The estimated total computational cost for each configuration is provided in table~\ref{tab: 1D flames}.

In order to adopt realistic initial flow conditions and inlet boundary conditions for the reactive simulations, a precursor simulation of a non-reactive, fully-developed TCF with streamwise and spanwise periodicity is conducted, in which the flow is driven by an imposed streamwise pressure gradient. The reactive simulations are initialized by superimposing the thermochemical state of homologous (nominal) 1D laminar flames, parametrized in an assumed V-shape, onto an instantaneous instance of the turbulent flow field provided by the precursor simulation. The time-dependent velocity fluctuations required at the upstream inlet boundary of the reactive simulations are imposed as temporally evolving turbulence, obtained by sampling in time a cross section of the inert precursor simulation. This procedure provides realistic turbulent inflow conditions at the inlet of the reactive simulations \citep{gruber_turbulent_2010,gruber_direct_2012,gruber_direct_2015,gruber_direct_2018}.
Instantaneous samples of the solution used for the analysis presented below are extracted after a sufficient time from the start of the simulations (beyond $2\,\ce{ms}$), once they have reached steady-state conditions.

\begin{figure}[ht!]
    \centering
    \includegraphics[width=0.75\linewidth]{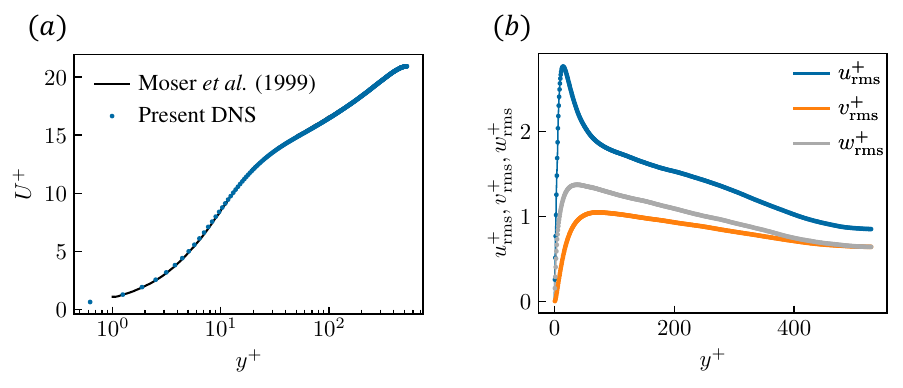}
    \caption{Comparison of the non-reactive, fully-developed turbulent channel flow precursor DNS at $Re_\tau = 530$ with DNS data from \citet{moser_direct_1999} at $Re_\tau = 590$. $(a)$ Non-dimensional mean streamwise velocity $U^+$. $(b)$ Non-dimensional streamwise rms velocity fluctuation $u^+_\mathrm{rms}$.}
    \label{fig: DNS flow vali}
\end{figure}
As the S3D solver has not previously been applied to predict TCF configurations at such an elevated friction Reynolds number ($Re_\tau = 530$), the non-reactive precursor DNS is used to validate the characteristic flow profiles against the DNS of \citet{moser_direct_1999} at $Re_\tau = 590$. Figure~\ref{fig: DNS flow vali}$(a)$ shows the normalized mean streamswise-velocity profile $U^+ = \langle u \rangle / U_\tau$ as a function of wall-normal distance $y^+$, while figure~\ref{fig: DNS flow vali}$(b)$ presents the corresponding normalized root mean square (rms) velocity fluctuations $u_\mathrm{rms}^+=\sqrt{\langle (u-U)^2 \rangle} / U_\tau$, where $u$ is the instantaneous, $U$ the averaged velocity field and $U_\tau$ the friction velocity \citep{Pope2000}. The results are in excellent agreement with the reference data, confirming the applicability of the flow solver at the Reynolds number considered in this study.


\section{Results} \label{sec: results}

The presentation of the results is structured as follows: first, the global propagation characteristics of the turbulent flames are examined in section~\ref{sec: global turbulent flame}; next, the local flame structure is analyzed in section~\ref{sec: local flame structure}; lastly, section~\ref{sec: near-wall propagation} investigates the near-wall flame propagation, including an analysis of the intermittent FWI/quenching process. 


\subsection{Global characteristics of the turbulent flames} \label{sec: global turbulent flame}

The DNS provides fully-resolved 3D instantaneous data of the flow field and of the thermochemical state.
The present flame configuration allows for spatial averaging along the statistically homogeneous spanwise direction ($z^+$) and in time. In addition, spatial averaging is performed across the symmetry plane $y^+=530$. 
A reaction progress variable $PV = Y_{\ce{H2O}} - Y_{\ce{H2}} - Y_{\ce{O2}}$ is used to define the flame front and is  normalized as a function of the local mixture composition:
\begin{equation} \label{eq: PV norm}
    C_\mathrm{norm} = \frac{PV - PV_\mathrm{min}(Z_\mathrm{Bilger})}{PV_\mathrm{max}(Z_\mathrm{Bilger})-PV_\mathrm{min}(Z_\mathrm{Bilger})}\,.
\end{equation} 
In equation~\ref{eq: PV norm}, $Z_\mathrm{Bilger}$ is the mixture fraction following the definition from \citet{BILGER1990135}, and $PV_\mathrm{min}$ and $PV_\mathrm{max}$ are obtained from a database of 1D laminar FP flames with equivalence ratio corresponding to the local mixture fraction.
This normalized progress variable definition accounts for local mixture variation due to differential and preferential diffusion effects, as well as super-adiabatic states resulting from TD phenomena. It enables a threshold selection for the flame iso-surface that consistently represents the flame surface along its wrinkled front, at both positively- and negatively-curved regions. This is a crucial feature required to properly and consistently analyze flames susceptible to TD phenomena \citep{howarth_empirical_2022, yao_isolating_2024}.

\begin{figure}[ht!]
    \centering
    \includegraphics[width=1.05\textwidth]{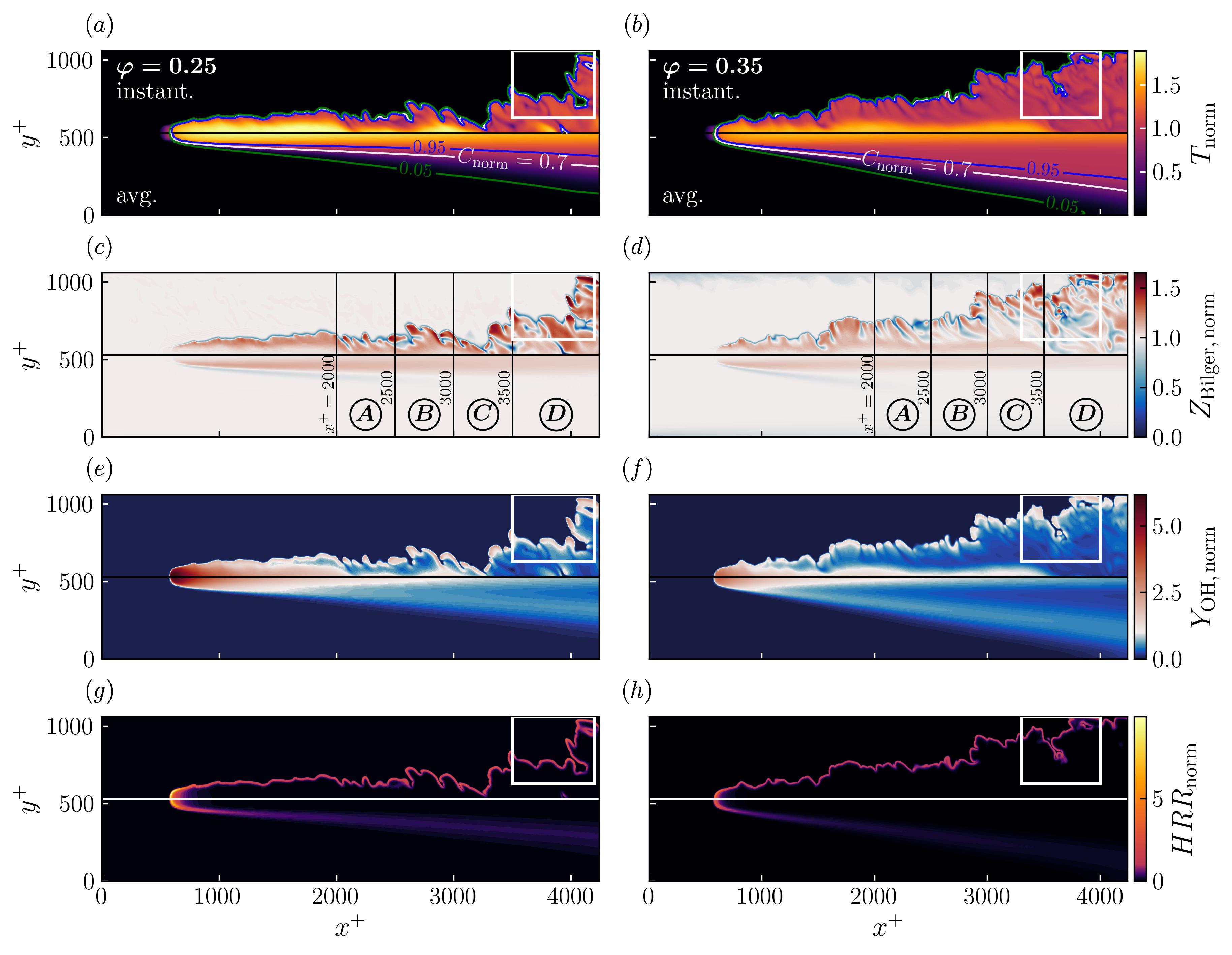}
    \caption{
    Instantaneous and averaged flame structures for both simulated flames $\varphi=0.25$ $(a,c,e,g)$ and $\varphi=0.35$ $(b,d,f,h)$. 
    Panels $(a,b)$ show the normalized fields of temperature $T_\mathrm{norm}$; $(c,d)$ the Bilger's mixture fraction $Z_\mathrm{Bilger,\,norm}$; $(e,f)$ the \ce{OH} mass fraction $Y_{\ce{OH},\,\mathrm{norm}}$; and $(g,h)$ the heat-release rate $HRR_\mathrm{norm}$ in physical space.
    The colored lines in the temperature field $(a,b)$ denote different iso-values of the normalized progress variable $C_\mathrm{norm}$, where $C_\mathrm{norm}=0.7$ is used to define the flame front.
    In panels $(c,d)$ the vertical lines mark the boundaries of the four spatial bins $A$-$D$, positioned at different longitudinal locations (in the streamwise direction), that will be used in the statistical analysis. White boxes highlight the region which is magnified and shown in figure \ref{fig: avg_instant_flame_zoom}.
    }
    \label{fig: avg_instant_flame}
\end{figure}
Figure~\ref{fig: avg_instant_flame} presents wall-to-wall longitudinal sections of the 3D domain for both flames: $\varphi = 0.25$ in $(a,c,e,g)$ and $\varphi = 0.35$ in $(b,d,f,h)$.
The normalized temperature $T_\mathrm{norm}$, mixture fraction $Z_\mathrm{Bilger,\,norm}$, \ce{OH} radical mass fraction $Y_{\ce{OH},\,\mathrm{norm}}$, and heat-release rate $HRR_\mathrm{norm}$ are shown as instantaneous (inst.) and averaged (avg.) fields in the upper and lower half of each panel, respectively.
All quantities are normalized by the homologous (nominal) 1D laminar FP flame values: the temperature and the mixture fraction are normalized by the adiabatic flame temperature and mixture fraction of the unburnt gas, respectively, while the \ce{OH} mass fraction and the heat-release rate are normalized using their respective peak values.
In figure \ref{fig: avg_instant_flame} $(a,b)$, iso-contours of the normalized progress variable are superimposed on the temperature field. In the following analysis, an iso-value of $C_\mathrm{norm} = 0.7$ is used to define the flame front, as it closely corresponds to the location of peak heat-release rate along the wrinkled flame surface.
Figure~\ref{fig: avg_instant_flame} shows that the $\varphi=0.25$ flame remains largely confined to the outer region of the core flow and does not penetrate into the near-wall region, whereas the $\varphi=0.35$ flame propagates much closer to the walls, encountering stronger turbulence and interacting with the wall further downstream. The instantaneous images reveal that, immediately downstream of the flame holder, both flames develop pronounced wrinkling and form the characteristic cellular structures associated with flames susceptible to TD phenomena. The curvature of these cellular structures, along with the overall wrinkling of the global flame front, intensifies as the flames propagate downstream and toward the walls, as qualitatively evidenced by the instantaneous snapshots of the solution.
Figure~\ref{fig: avg_instant_flame}$(c,d)$ clearly highlights the variation of the local mixture fraction along the flame front, with locally enriched mixture in the positively-curved cellular domes, separated by narrow cusps characterized by locally leaner mixture.
The instantaneous fields of $Y_{\ce{OH},\,\mathrm{norm}}$ and $HRR_\mathrm{norm}$ in figure~\ref{fig: avg_instant_flame}$(e$-$h)$ exhibit significant variation along the flame front, highlighting positively-curved cellular domes and structures with locally enriched mixture and higher reactivity, separated by negatively-curved narrow cusps with lower reactivity.
In both flames ($\varphi=0.25$ and $\varphi=0.35$), typical effects of TD phenomena are present: super-adiabatic temperatures that exceed the nominal adiabatic flame temperature ($T_\mathrm{norm}>1$), local mixture enrichment ($Z_\mathrm{Bilger,\,norm}>1$), formation of cellular structures and strong spatial variations in reactivity indicators such as the \ce{OH} radical and heat-release rate.

\begin{figure}[ht!]
    \centering
    \includegraphics[width=0.9\textwidth]{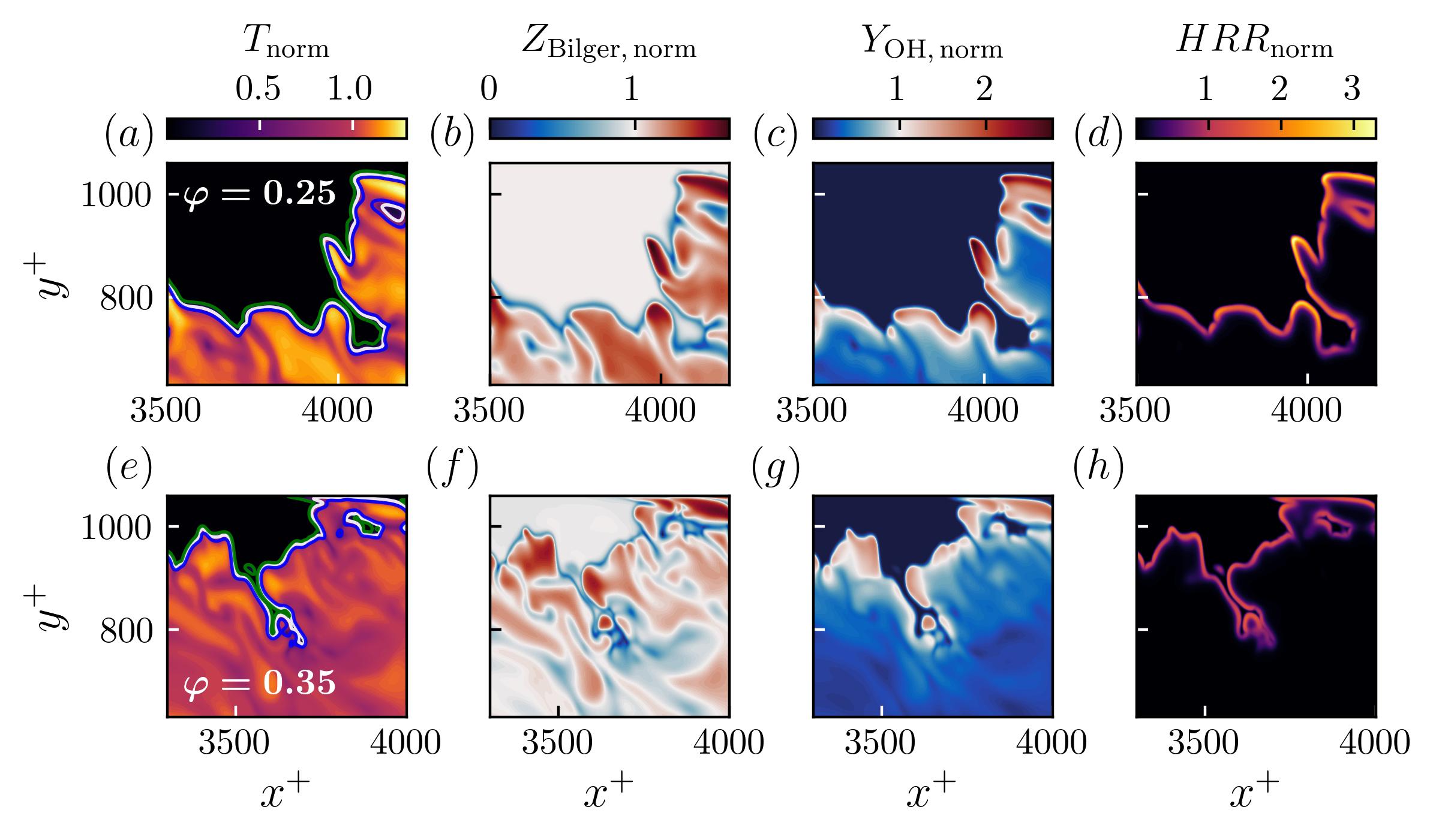}
    \caption{Zoomed view of instantaneous flame structures corresponding to the white boxes in figure \ref{fig: avg_instant_flame} for the $\varphi=0.25$ flame $(a$-$d)$ and the $\varphi=0.35$ flame $(e$-$h)$.}
    \label{fig: avg_instant_flame_zoom}
\end{figure}
To gain further insights into the local flame structure, figure~\ref{fig: avg_instant_flame_zoom} presents a magnified view of the regions highlighted by the white boxes in figure \ref{fig: avg_instant_flame}. Close inspection of the instantaneous flame front reveals the characteristic local flame response associated with TD phenomena. Positively-curved bulges (convex toward the unburned mixture) display super-adiabatic temperatures ($T_\mathrm{norm}>1$), locally enriched mixture fractions ($Z_\mathrm{Bilger,\,norm} > 1$), elevated radical concentrations, and intensified heat-release rates. In contrast, negatively-curved cusps exhibit locally leaner mixtures with reduced temperature and reactivity. The reactivity variations present along the wrinkled flame front, accelerating some regions while decelerating others, lead to flame stretch and substantial increase in flame surface area, ultimately enhancing the overall burning rate of the turbulent flame.
As expected, visual comparison of the normalized fields shows that the $\varphi = 0.25$ flame (figure~\ref{fig: avg_instant_flame_zoom}$a$–$d$) attains higher peak values than the $\varphi = 0.35$ flame (figure~\ref{fig: avg_instant_flame_zoom}$e$–$h$). Previous laminar studies have demonstrated that the intensity of TD phenomena increases with decreasing equivalence ratio \citep{Frouzakis2015, Berger2022_par1, Berger2022_part2}. The present results confirm that this trend persists for TD-susceptible flames interacting with near-wall shear turbulence.


\subsubsection{Regime diagram classification}

To broadly characterize the turbulent flame behavior, the combustion regime must be determined first, as it identifies the relationship between turbulent and chemical scales. This classification is essential for practical applications, since many turbulent combustion processes and models are calibrated for specific regime conditions. The Karlovitz number ($Ka$) is a key dimensionless parameter used to quantify the local interaction between the smallest turbulent scales and the chemical scales of the flame \citep{williams_combustion_1985, Peters2000}. In this work, $Ka$ is evaluated following the procedure of \citet{gruber_direct_2018} for reactive TCF configurations. It is defined as the ratio of the local flame time to the Kolmogorov time scale, $Ka = t_\mathrm{flame}/t_\eta$. In the present TCF configuration, the (local) Kolmogorov time scale is a function of the wall distance, its value decreasing from the core flow towards the near-wall region of the flow, and it is obtained from the DNS flow field as $t_\eta = (\nu/\varepsilon)^{1/2}$, where $\nu$ denotes the kinematic viscosity and $\varepsilon$ the turbulent dissipation rate \citep{Pope2000}. 
A flamelet table is generated from 1D FP flames spanning varying mixture compositions, which is used to determine the local flame time scale $\tau_{_\mathrm{flame}} = \delta_\mathrm{F} / s_\mathrm{L}$ in the turbulent flame, as a function of the local mixture composition $Z_\mathrm{Bilger}$ and reaction progress $C_\mathrm{norm}$.

\begin{figure}[ht!]
    \centering
    \includegraphics[width=0.55\linewidth]{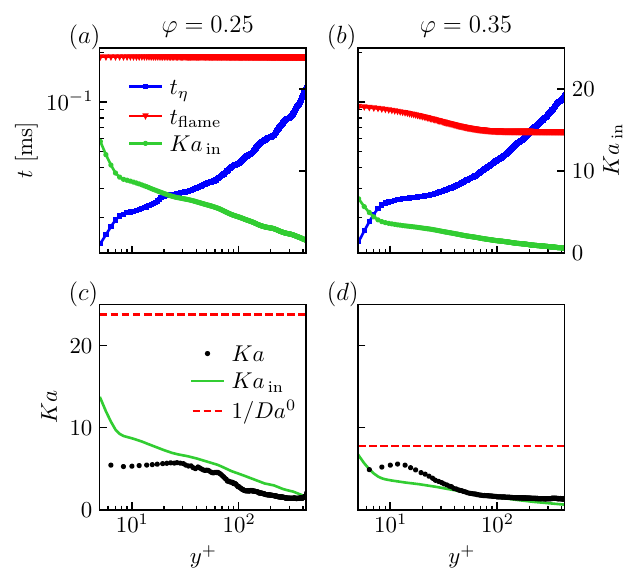}
    \caption{Karlovitz number as a function of the wall-normal distance $y^+$. $(a, b)$ show the local flame time $t_\mathrm{flame}$ and Kolmogorov timescale $t_\eta$, as well as the resulting Karlovitz number $Ka_\mathrm{in}$ in correspondence of the inlet plane ($x^+=0$) for the $\varphi=0.25$ and the $\varphi=0.35$ flame, respectively. $(c, d)$ show the effective Karlovitz number $Ka$ immediately upstream the flame front ($0.1 \leq C_\mathrm{norm} \leq 0.3$) for both flames. The Karlovitz number at the inlet and the inverse of the nominal Damköhler number are shown for comparison.}
    \label{fig: Karlovitz}
\end{figure}
Figure~\ref{fig: Karlovitz}$(a, b)$ illustrates the wall-normal profiles, in correspondence of the domain inlet ($x^+=0$), of the turbulent ($t_\eta$) and chemical ($t_\mathrm{flame}$) time scales, and of the resulting \textit{inlet} Karlovitz number ($Ka_\mathrm{in}$) for both flames simulated. The channel wall is located at $y^+=0$ in the figures.
Because the lower equivalence ratio yields a larger chemical time scale $t_\mathrm{flame}$, the $\varphi = 0.25$ flame exhibits higher inlet Karlovitz numbers $Ka_\mathrm{in}$ than the $\varphi = 0.35$ flame for identical inlet turbulent time scales $t_\eta$ (figure~\ref{fig: Karlovitz}$a,b$). The effective Karlovitz number $Ka$, evaluated immediately upstream of the flame front ($0.1 \leq C_\mathrm{norm} \leq 0.3$), is shown in figure~\ref{fig: Karlovitz}$c,d$. For both equivalence ratios, $Ka$ increases monotonically from the core region toward the walls. In the $\varphi = 0.35$ case (figure~\ref{fig: Karlovitz}$d$), a pronounced local maximum appears, coinciding with the peak in turbulence intensity within the buffer layer; further discussion of this behavior is provided in section~\ref{sec: near-wall propagation}. For reference, figure~\ref{fig: Karlovitz} also reports the inlet Karlovitz number $Ka_\mathrm{in}$ and the inverse nominal Damköhler number $Da^0$ (table~\ref{tab: 1D flames}). Consistent with the observations of \citet{gruber_direct_2018}, a single global parameter, such as the nominal Damköhler number, cannot adequately characterize the combustion regime in the TCF configuration, as both turbulent and chemical time scales vary substantially with wall distance. As previously argued \citep{gruber_direct_2018}, the inlet profile of $Ka_\mathrm{in}$ provides a more representative approximation of the effective Karlovitz number governing flame behavior. However, because the anchored V-shaped flame significantly alters the local flow field in the downstream region of the flow and the presence of the wall induces significant heat loss, the correspondence between $Ka_\mathrm{in}$ and the effective $Ka$ breaks down in the near-wall region.
%
For the $\varphi = 0.25$ flame, the results in figure~\ref{fig: Karlovitz}$(c)$ should be interpreted with caution, as only a limited number of flame realizations reach the near-wall region ($y^+ \lesssim 100$). It is also important to note that the Karlovitz numbers reported in figure~\ref{fig: Karlovitz}$(c,d)$ are computed using local values of the flame time, leading to an approximately five-fold increase in $Ka$ between the core flow and the wall. If, instead, the nominal values of the flame time are used, the near-wall rise in Karlovitz number is even more pronounced, approximately 16-fold for the $\varphi = 0.25$ flame and 8-fold for the $\varphi = 0.35$ flame.

Based on the analysis of the Karlovitz number, the simulated flames are classified according to their combustion regime. The combustion regime is uniquely defined by two of the following quantities \citep{aspden_turbulenceflame_2011}: the flame Reynolds number $Re_\mathrm{F}=u_\mathrm{rms}\ell_0 \,/\, (s_\mathrm{L} \delta_\mathrm{F})$, the Karlovitz number $Ka^2 = u_\mathrm{rms}^3 \delta_\mathrm{F} \,/\, (s_\mathrm{L}^3 \ell_0)$ and the Damköhler number $\mathrm{Da}=s_\mathrm{L} \ell_0 \,/\, (u_\mathrm{rms} \delta_\mathrm{F})$, where $\ell_0$ is the integral length scale of the flow.
\begin{figure}[ht!]
    \centering
    \includegraphics[width=0.55\textwidth]{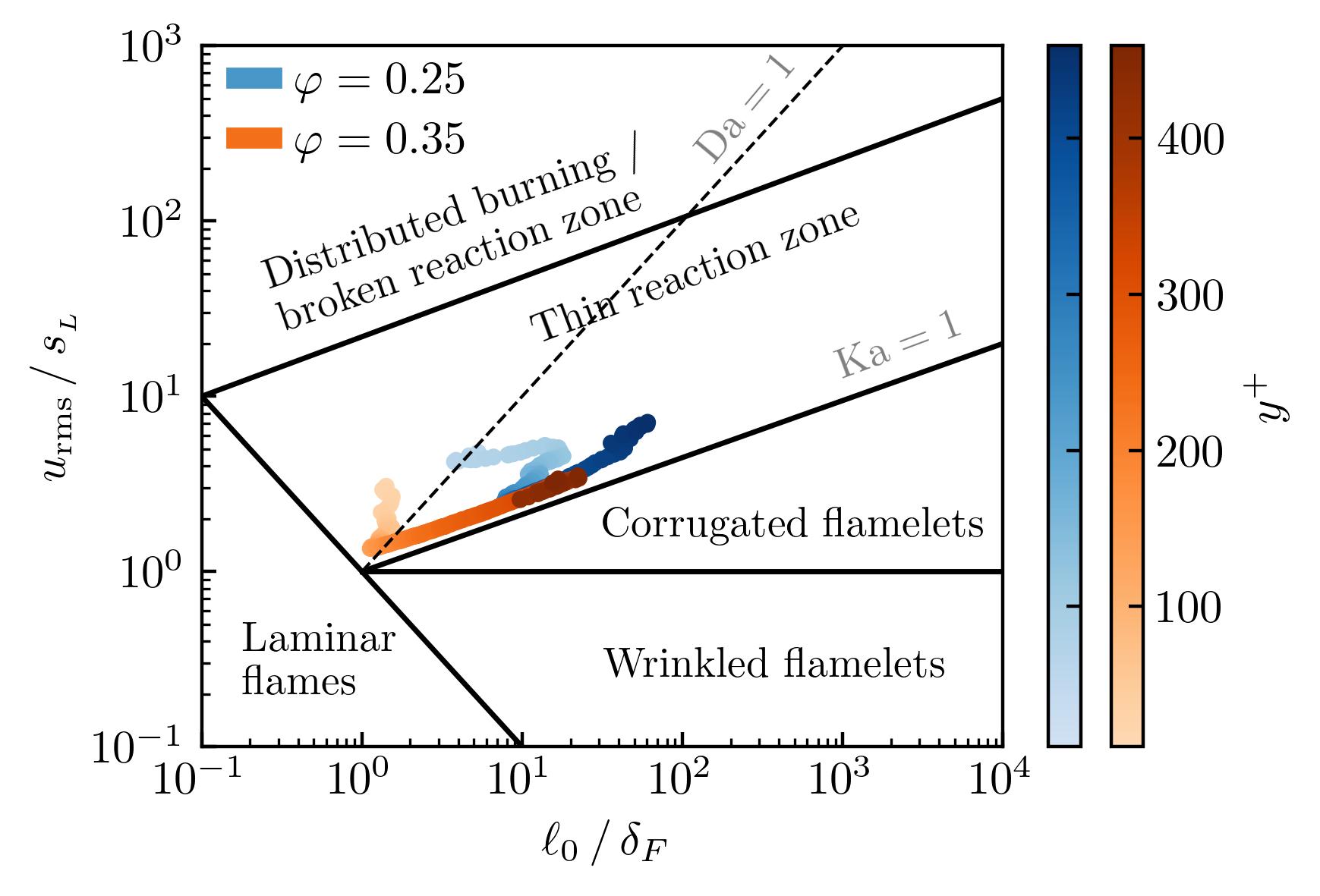}
    \caption{Peters-Borghi Regime Diagram of the flame front at different wall-normal positions for both flames $\varphi=0.25$, $\varphi=0.35$.}
    \label{fig: regime diagram}
\end{figure}
Figure~\ref{fig: regime diagram} positions the simulated flames on the Peters–Borghi diagram \citep{PETERS1999} as a function of the wall-normal distance $y^+$. Both flames reside within the thin reaction zone regime, characterized by Karlovitz numbers greater than unity and Kolmogorov scales smaller than the laminar flame thickness. As the wall is approached, the increasing turbulence intensity shifts the flames toward the distributed burning regime. 


\subsubsection{Turbulent flame speed and flame surface area}

In turbulent combustion modeling and for accurate numerical prediction of combustion processes, the turbulent flame speed and the flame surface area are of fundamental interest.
Based on the early work of \citet{Damkhler1940}, the modified first Damköhler hypothesis has emerged, which is relevant for many combustion models and postulates a direct relation between these quantities for turbulent flame surface wrinkling:
\begin{equation} \label{eq: I0}
    \frac{s_\mathrm{T}}{s_\mathrm{L}} = I_0 \frac{A_\mathrm{T}}{A_0}\,.
\end{equation}
In equation \ref{eq: I0}, $s_\mathrm{T}$ is the turbulent flame speed, $A_\mathrm{T}$ is the instantaneous turbulent flame surface area, $A_0$ is the fuel cross-sectional area and $I_0$ is known as the stretch factor \citep{Bray_1990, Bray_1991}, and also referred to in the literature as a reactivity factor \citep{PartI}, burning factor \citep{Bradley2011} or burning efficiency \citep{yao_isolating_2024}.

For turbulent combustion of conventional hydrocarbon fuels in air, where the effective Lewis number is close to unity, the first Damköhler hypothesis holds and the stretch factor remains near unity ($I_0 \sim 1$) \citep{Nivarti2017}. In contrast, flames with effective Lewis numbers well below unity are susceptible to TD phenomena, which can substantially elevate local reactivity and burning rate. These localized enhancements accelerate the global flame front beyond what would be expected from surface-area wrinkling alone, leading to stretch-factor values that may exceed unity by several fold.

In TD-susceptible premixed flames, the stretch factor provides a quantitative measure of their departure from the conventional behavior described by the first Damköhler hypothesis, and thus reflects the relative influence of TD effects on the burning rate. Several recent studies \citep{howarth_empirical_2022, rieth_effect_2023, berger_combustion_2025, Schneider2025} suggest that accurate prediction of the stretch factor is essential for reliably capturing the propagation characteristics of hydrogen flames affected by TD phenomena.

In the following analysis, the turbulent flame speed is estimated by evaluating the consumption speed, computed from the volume-integrated hydrogen source term in each one of the longitudinal bins $A$-$D$:
\begin{equation} \label{eq: sC}
    s_\mathrm{T}= - \frac{1}{\rho_\mathrm{u} A_0 Y_{\ce{H2}, \,\mathrm{u}}} \int_{_V} \dot{\omega}_{\ce{H2}} \,\mathrm{d}V\,.
\end{equation}
In equation \ref{eq: sC}, $\rho_\mathrm{u}$ and $Y_{\ce{H2}, \,\mathrm{u}}$ denote the density and mass fraction of hydrogen in the unburnt mixture, and $A_0$ is the mean flame-surface area present within each longitudinal bin. The hydrogen source term $\dot{\omega}_{\ce{H2}}$ is integrated within the volume $V$ of the corresponding bin. Therein, the instantaneous, turbulent flame surface area $A_\mathrm{T}$ is evaluated geometrically from the interpolated $C_\mathrm{norm}=0.7$ iso-surface using triangulation \citep{sullivan2019pyvista}.

\begin{figure}[ht!]
    \centering
    \includegraphics[width=0.6\linewidth]{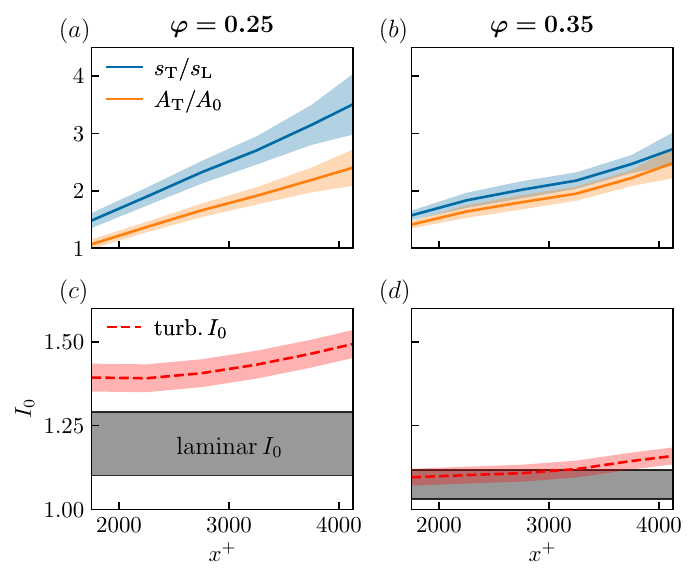}
    \caption{Spatial evolution in the streamwise direction of the normalized turbulent flame speed ($s_\mathrm{T}/s_\mathrm{L}$) and of the surface-area ratio ($A_\mathrm{T}/A_0$) at $\varphi=0.25$ in $(a)$ and $\varphi=0.35$ in $(b)$. The colored, shaded corridors indicate one standard deviation around the mean. The stretch factor $I_0$ is shown for both turbulent flames in $(c, d)$, respectively. The gray-shaded area represents the range of $I_0$ values at laminar reference conditions.}
    \label{fig: sC AT I0 xbins}
\end{figure}
Figure~\ref{fig: sC AT I0 xbins}$(a,b)$ shows the longitudinal evolution along the streamwise direction of the turbulent flame speed, normalized by the respective laminar burning velocity, and of the instantaneous flame-surface area, normalized by the respective mean flame-surface area, for the $\varphi=0.25$ and $\varphi=0.35$ flames, respectively. Both quantities increase for increasing longitudinal (streamwise) position, fluctuations are highlighted by the colored corridors that mark one standard deviation around the mean. Notably, the leaner flame at $\varphi=0.25$ exhibits a greater enhancement in turbulent flame speed, while both flames yield a similar increase in surface area.
Figure~\ref{fig: sC AT I0 xbins}$(c,d)$ illustrates the longitudinal evolution in the streamwise direction of the stretch factor $I_0$. 
For reference, the range of $I_0$ values that is found in (initially planar) 2D laminar flames susceptible to TD phenomena is shown by the gray corridor. It is important to note that the stretch factor estimated from these 2D laminar configurations strongly depends upon the domain size in the transversal direction and, thereby, upon the resulting geometrical confinement of certain wavelengths, as demonstrated by \citet{berger_characteristic_2019}. This observation is also reproduced for the present operating conditions in section~S1.2 of the supplementary document. 

Crucially, in turbulent flames, the growth of flame-surface instabilities induced by TD phenomena is modulated by the characteristic time and length scales of the surrounding fluid motions, which may either suppress or amplify the instability, rather than by the size of the computational domain.

Figure \ref{fig: sC AT I0 xbins}$(c,d)$ shows a pronounced rise in the stretch factor for both turbulent flames at downstream locations beyond $x^+ \approx 3000$, indicating significant synergistic interactions between turbulence and TD phenomena. As expected, the leaner $\varphi = 0.25$ flame, for which TD effects are more pronounced, exhibits the larger values of $I_0$. For the $\varphi = 0.35$ flame, the stretch factor exceeds its laminar reference only for $x^+ \gtrsim 3300$, coinciding with the region where the flame approaches the wall and enters the buffer layer, characterized by elevated turbulence intensity. Section \ref{sec: near-wall propagation} provides further evidence that increases in $I_0$ correlate with heightened levels of turbulence intensity.


To quantitatively assess the differences between the two flames, the magnitude of the progress-variable gradient $|\nabla PV|$, the displacement speed $s_d$, and the flame stretch $K$ are evaluated at the flame front ($C_\mathrm{norm} \approx 0.7$). The displacement speed of the iso-surface describing the flame front is computed following the methodology of \citet{Echekki1999}, based on a decomposition into reaction, diffusion, and curvature contributions. The density-weighted displacement speed is defined as $\widetilde{s}_\mathrm{d} = (\rho s_\mathrm{d}) / \rho_\mathrm{u}$. Flame stretch is given by $K = K_\mathrm{s} + s_\mathrm{d},\kappa_\mathrm{c}$, where $K_\mathrm{s}$ is the strain-rate contribution and $\kappa_\mathrm{c}$ is the flame curvature \citep{MATALON1983, Altantzis2012}. The strain rate $K_\mathrm{s} = \nabla_t \cdot \mathbf{u}_t$ is obtained from the divergence of the tangential velocity, and the mean flame curvature $\kappa_\mathrm{c} = \nabla \cdot \mathbf{n}$ is evaluated from the flame-normal vector $\mathbf{n} = -\nabla PV / |\nabla PV|$, pointing toward the unburned mixture.

\begin{figure}[ht!]
    \centering
    \includegraphics[width=0.725\linewidth]{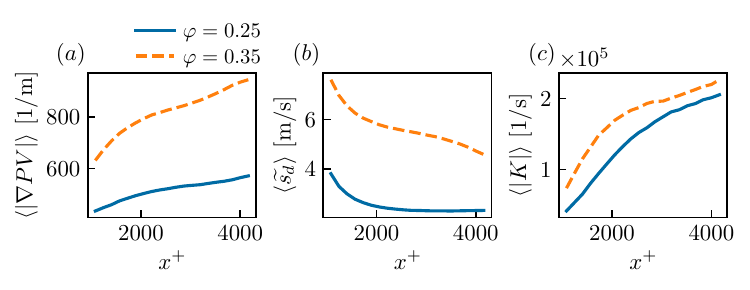}
    \caption{Spatial evolution in the streamwise direction of the mean values of $(a)$ the absolute progress-variable gradient $|\nabla PV|$, $(b)$ the density-weighted displacement speed $\widetilde{s}_\mathrm{d}$, and $(c)$ the absolute flame stretch rate $|K|$. All quantities are evaluated at the flame front ($C_\mathrm{norm}\approx0.7$).
    }
    \label{fig: displacement speed}
\end{figure}
Figure~\ref{fig: displacement speed} shows the streamwise evolution of the mean absolute progress-variable gradient $(a)$, mean displacement speed $(b)$, and mean absolute flame-stretch rate $(c)$. With increasing streamwise position, both flames display the same overall behavior: flame stretch increases, the local flame thickness decreases (reflected in larger progress-variable gradients), and the displacement speed gradually declines.

The mean displacement speed reaches $1.72$ times the laminar burning velocity for the $\varphi = 0.35$ flame and $1.84$ times for the $\varphi = 0.25$ flame. The consistently higher values of $|\nabla PV|$, $\widetilde{s}_\mathrm{d}$, and $|K|$ for $\varphi = 0.35$ indicate a locally thinner and faster flame that is subjected to stronger stretch as it propagates closer to the walls, where turbulence intensity is elevated.


\subsection{Local analysis of the topology and microstructure of the turbulent flame} \label{sec: local flame structure}

First, the topology and curvature radius of the flame front are examined. The local flame microstructure is then analyzed, with emphasis on flame curvature, as it plays a critical role in modulating the effects of TD phenomena on the flame-front acceleration \citep{Rocco2014, howarth_empirical_2022, bottler_can_2024}.


\subsubsection{Local flame topology and most probable curvature}

To quantify the geometrical features of the flame front, the principal curvatures $\kappa_1$ and $\kappa_2$ are computed as the minimum and maximum eigenvalues of the gradient of the flame normal vector \citep{cecere_direct_2016}. They are related to the mean curvature $\kappa_\mathrm{c} = \nabla \cdot \mathbf{n} = (\kappa_1 + \kappa_2)/2$ and the Gaussian curvature $\kappa_\mathrm{g} = \kappa_1 \, \kappa_2$ \citep{Han2019}. Based on the sign and magnitudes of the principal curvatures (normalized by the respective thermal flame thickness $\delta_\mathrm{F}$), the flame surface topology can be classified into six canonical types \citep{howarth_thermodiffusively-unstable_2023, Wang2024}: flat flame (FF), leading point (LP), leading edge (LE), saddle point (SP), trailing edge (TE) and trailing point (TP).
Hereby, 'leading' and 'trailing' correspond to positive and negative curvature, respectively, and points and edges can be interpreted as spherically and cylindrical curved, respectively.
\begin{figure}[ht!]
    \centering
    \includegraphics[width=0.7\linewidth]{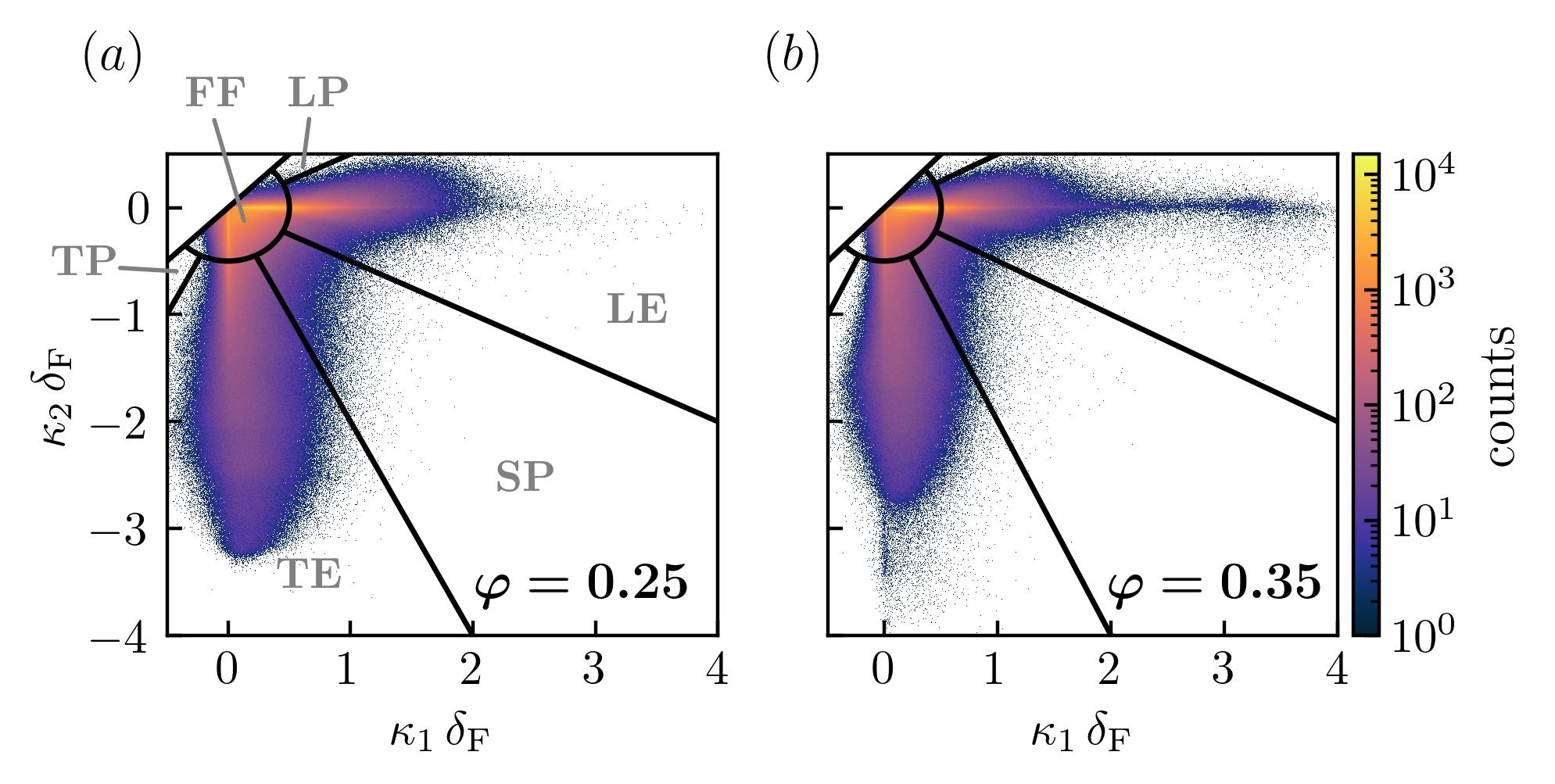}
    \caption{
    The joint PDFs of the principal curvatures $\kappa_1$ and $\kappa_2$, normalized by the respective laminar flame thickness and topologically classified into six zones, are shown for the $\varphi = 0.25$ flame in $(a)$ and the $\varphi = 0.35$ flame in $(b)$.}
    \label{fig: princ curvature distribution}
\end{figure}
Figure~\ref{fig: princ curvature distribution} shows the distribution of the flame surface classification for both $\varphi=0.25$ and $\varphi=0.35$ in $(a,b)$ flames, respectively. The analysis suggests that most of the flame surface is either flat (FF) or exhibits cylindrical curvature (LE, TE), characterized by the dominance of a single principal curvature. In the case of the richer $\varphi=0.35$ mixture, the distribution ranges to higher flame-curvature values. This is caused by the fact that the richer flame, more than the leaner one, propagates towards the wall and into regions of strong turbulent motions, leading to increased flame-surface deformation.

\begin{figure}[ht!]
    \centering
    \includegraphics[width=0.5\linewidth]{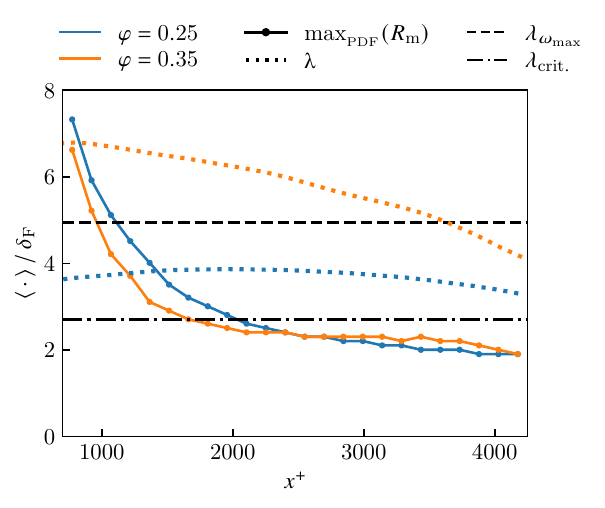}
    \caption{Spatial evolution in the streamwise direction of the most probable curvature radius $\max_{_\mathrm{PDF}}(R_\mathrm{m})$ of the flame front, normalized by the respective thermal flame thickness. For comparison, the most unstable wavelength $\lambda_{\omega_{\max}}$ and the critical wavelength $\lambda_\mathrm{crit.}$ are marked as horizontal lines while the Taylor microscale $\uplambda$ is shown by the colored dotted lines.
}
    \label{fig: mean curvature radius}
\end{figure}
Next, the curvature radius $R_\mathrm{m} = 1/|\kappa_\mathrm{c}|$ of the flame front is analyzed. This is related to the mean size of the cellular structures. Figure~\ref{fig: mean curvature radius} illustrates the spatial evolution in the streamwise direction of the most probable curvature radius of the flame front. For both flames, the most probable curvature radii show a rapid 3-fold decrease between the most upstream region of the flow considered and a longitudinal position of $x^+\approx2000$, a slower decrease is observed further downstream. This suggests that the most significant topological changes in the flame local morphology, mainly towards smaller structures characterized by stronger curvature, occur upstream of $x^+ \approx 2000$ while, downstream of this position, the size of the cellular structures remains nearly constant on average.
To compare the most probable curvature radii observed in the turbulent flames with the characteristic length scales of the intrinsic TD instabilities in the corresponding laminar flames, the most unstable wavelength $\lambda_{\omega_{\max}}$ and the critical wavelength $\lambda_\mathrm{crit}$ are shown as horizontal reference lines in figure~\ref{fig: mean curvature radius}\footnote{Note that $\lambda_{\omega_{\max}}$ and $\lambda_{\mathrm{crit}}$ are shown only for the $\varphi = 0.35$ mixture, as the corresponding values for both equivalence ratios are very close to one another.}. Both quantities are obtained from a numerical dispersion relation, following the procedure described in Section~S1.1 of the supplementary material. The most unstable wavelength $\lambda_{\omega_{\max}}$ represents the perturbation size associated with the maximum linear growth rate, whereas the critical wavelength $\lambda_\mathrm{crit}$ denotes the stability limit below which perturbations are attenuated.
The most probable curvature radius of both flames matches the most unstable wavelength at upstream locations of $x^+ \approx 1000$, suggesting that intrinsic flame TD instabilities dominate the flame topology in this region. However, further downstream (beyond $x^+\approx2000$), values of the curvature radii are close to the critical wavelength and become even smaller. This is an interesting observation, indicating that, in TD-susceptible flames that propagate in realistic and strongly anisotropic shear turbulence, flame structures smaller than the geometrical stability limit of TD instabilities can locally occur due to the non-linear flame response to turbulent forcing and modulation. Moreover, the present analysis also suggest that the Taylor microscale $\uplambda$, estimated from the DNS dataset and marked by the dotted lines in figure~\ref{fig: mean curvature radius}, is of the same order as the most probable curvature radii and characteristic length scales of intrinsic TD instabilities.


\subsubsection{Local flame microstructure}

\begin{figure}[ht!]
    \centering
    \includegraphics[width=0.825\linewidth]{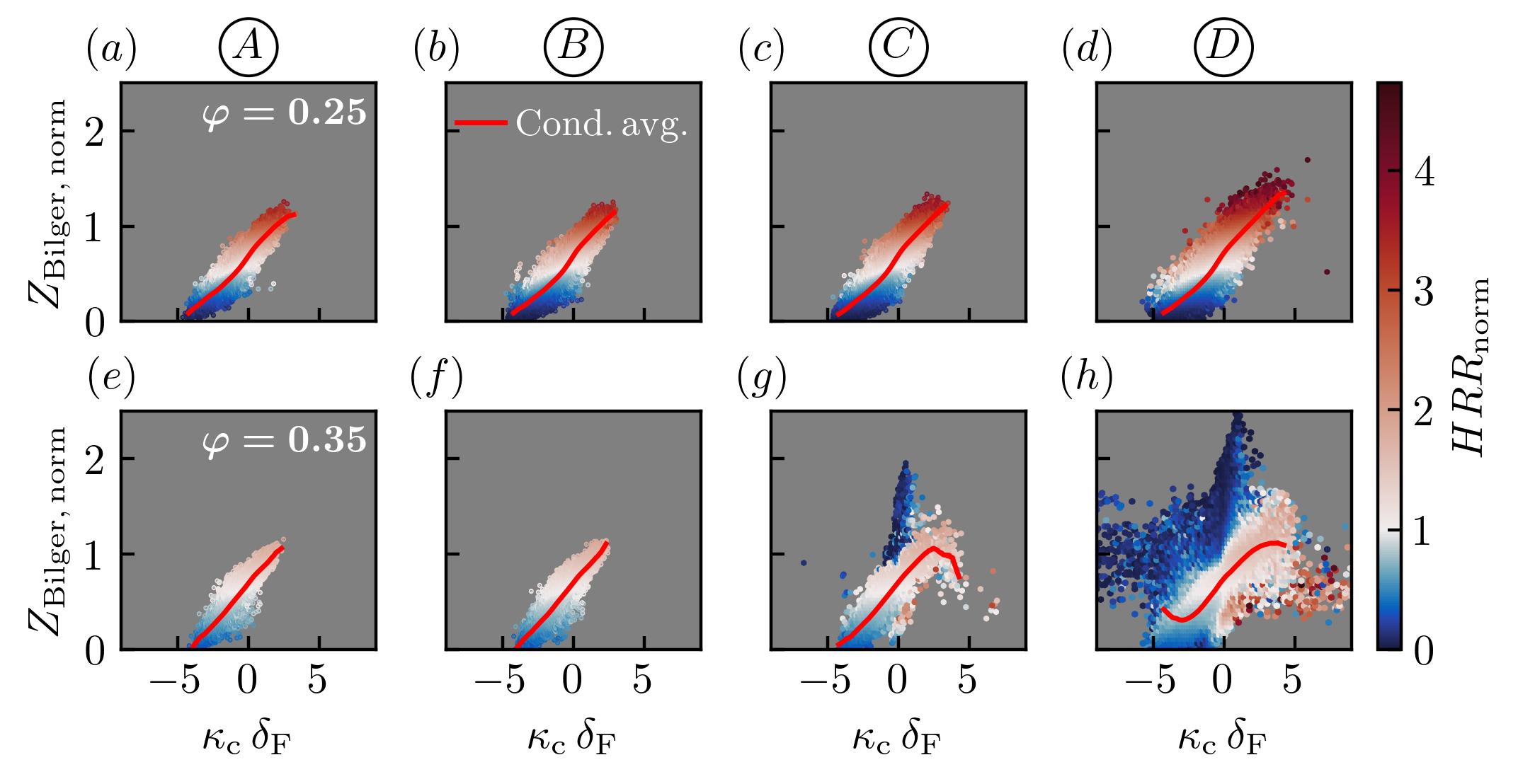}
    \caption{Normalized mixture fraction plotted against curvature of the flame front, colored by the normalized heat-release rate, in bins $A$–$D$ for flames $\varphi=0.25$ in $(a$-$d)$ and $\varphi=0.35$ in $(e$-$h)$. The red curve denotes the conditional average $\langle Z_\mathrm{Bilger,\,norm} \,|\, \kappa_\mathrm{c} \delta_\mathrm{F} \rangle$.}
    \label{fig: flame front ZBilger curv HRR}
\end{figure}
Figure~\ref{fig: flame front ZBilger curv HRR} analyzes the correlation between flame curvature and TD phenomena, showing the distribution of the normalized mixture fraction over curvature, colored by the normalized heat-release rate, for the flame front in correspondence of the four bins $A$–$D$ (cf. figure~\ref{fig: avg_instant_flame}). Consistent with previous observations, TD phenomena are more pronounced at the lower equivalence ratio ($\varphi=0.25$) with larger peak values of the normalized mixture fraction and heat-release rate. More specifically, the leaner flame at $\varphi=0.25$ (figure~\ref{fig: flame front ZBilger curv HRR}$(a$-$d)$) exhibits similar trends at all locations $A$–$D$: the mixture fraction increases in positively curved regions and decreases in negatively curved regions, revealing a strong correlation, which is also highlighted by the conditional average. This behavior is also reflected by the local heat-release rate variation, resulting in increased reactivity of positively curved regions. Also, peak values slightly increase with the streamwise position, suggesting a role of turbulence modulation in the characteristic response of the flame to TD phenomena. In the flame at $\varphi=0.35$ (figure~\ref{fig: flame front ZBilger curv HRR}$(e$-$h)$), similar trends apply to bins $A$ and $B$, while the flame structure differs in bins $C$ and $D$. This is due to flame propagation taking place closer to the walls, where the strongest turbulence intensity is present and FWI/quenching intermittently occurs. This is discussed in section~\ref{sec: near-wall propagation}.

\begin{figure}[ht!]
    \centering
    \includegraphics[width=0.825\linewidth]{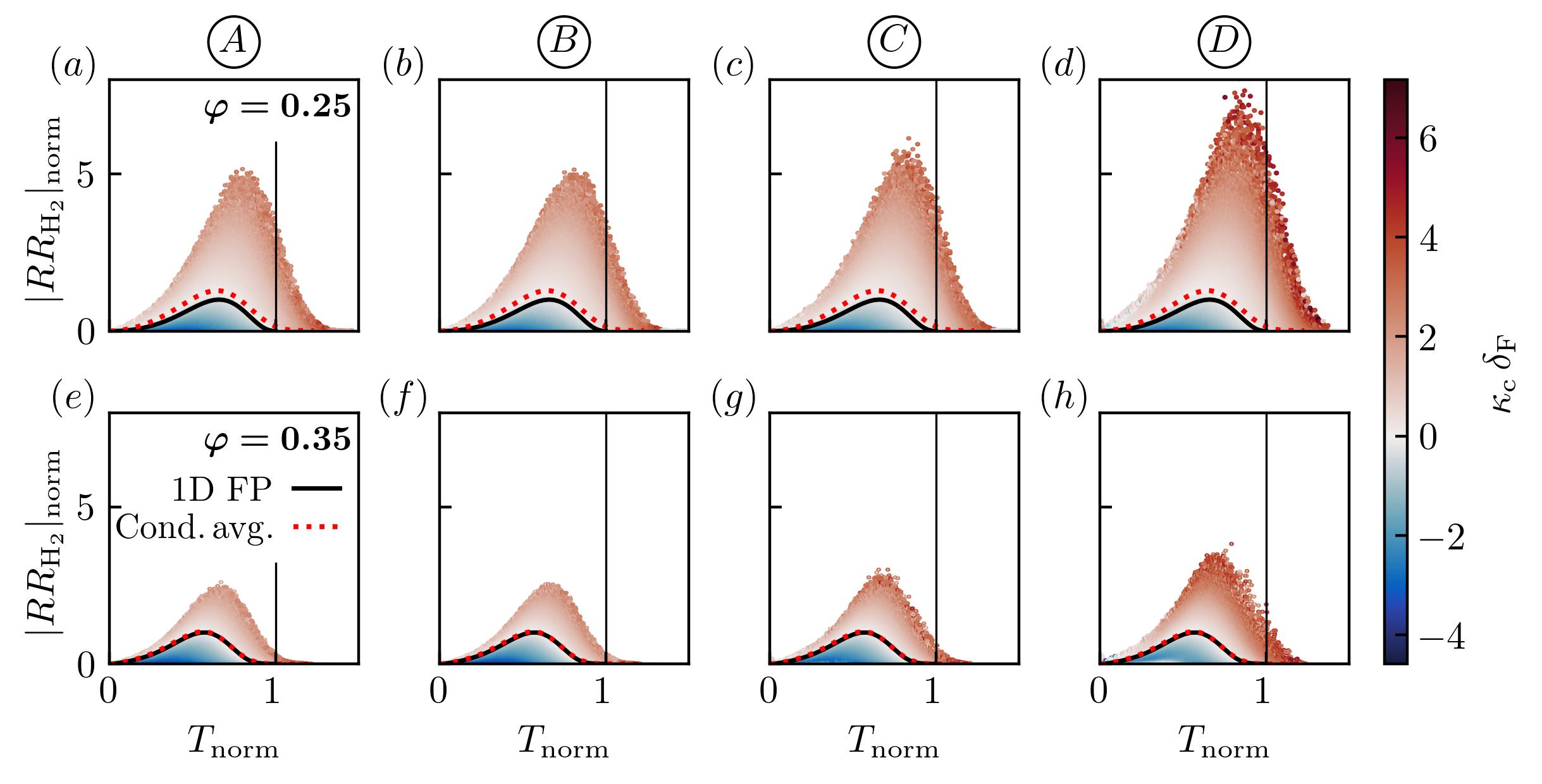
    }
    \caption{Thermochemical state space of the normalized reaction rate of hydrogen $|RR_{\ce{H2}}|_\mathrm{norm}$ versus the normalized temperature $T_\mathrm{norm}$, colored by flame curvature, in bins $A$–$D$ for both flames at $\varphi=0.25$ $(a$-$d)$ and $\varphi=0.35$ $(e$-$h)$. The dotted red line shows the conditional average $\langle |RR_{\ce{H2}}|_\mathrm{norm} \,\big|\, T_\mathrm{norm} \rangle$ while the black line represents the homologous (unstrained) 1D laminar FP flames. A vertical black line is placed in correspondence of the adiabatic flame temperature ($T_\mathrm{norm}=1$) to aid visualization of super-adiabatic states $T_\mathrm{norm}>1$.}
    \label{fig: state space RR_H2 Tnorm}
\end{figure}
To quantify the correlation between flame curvature, local reactivity and super-adiabatic temperature, figure~\ref{fig: state space RR_H2 Tnorm} presents the state space of the normalized absolute reaction rate of hydrogen $|RR_{\ce{H2}}|_\mathrm{norm}$ over the normalized temperature $T_\mathrm{norm}$, colored by the local flame curvature. The normalized absolute reaction rate of hydrogen $|RR_{\ce{H2}}|_\mathrm{norm}$ serves as a measurement of overall reactivity. The DNS data clearly evidences that higher magnitudes of normalized reaction rate and normalized temperature are found in the $\varphi=0.25$ flame (figure~\ref{fig: state space RR_H2 Tnorm}$(a$-$d)$).
Furthermore, Figure~\ref{fig: state space RR_H2 Tnorm} reveals the following trends for both flames:
\begin{itemize}
\item States with increased reaction rate or super-adiabatic temperature ($T_\mathrm{norm}>1$) exhibit positive curvature. 
\item Super-adiabatic temperatures occur mainly behind the peak reaction rate in the post-oxidation zone and they are governed by high positive curvature and therefore enriched mixture.
\item Further downstream, the values of curvature, reaction rate and flame temperature increase, indicating amplification of the TD phenomena with increasing turbulence intensity modulating the response of the flame.
\end{itemize}
%
For reference, the homologous (unstrained) 1D laminar FP flames are shown by the black curves, aligning well with flat and mildly-curved portions of the flame front and separating flame structures subject to positive and negative curvature.
Notably, the conditional average (red dotted lines) closely matches the unstrained 1D laminar FP flame solution at $\varphi = 0.35$ (figure~\ref{fig: state space RR_H2 Tnorm}~$(e$-$h)$), while it clearly exceeds it at $\varphi=0.25$ (figure~\ref{fig: state space RR_H2 Tnorm}$(a$-$d)$), highlighting more pronounced TD phenomena acting on the flame structure.


\subsection{Flame propagation in the vicinity of the wall} \label{sec: near-wall propagation}

The core flow features relatively large scales and low turbulence intensity, leading to a qualitatively similar flame response at both equivalence ratios investigated. However, as the flame front propagates closer to the wall, it encounters varying local conditions and an increasing level of turbulence intensity. Therefore, a variation in the flame response is expected between the core flow and the near-wall region of the flow, with flame quenching intermittently occurring at the ‘‘cold’’ isothermal walls.
As reported by \citet{gruber_turbulent_2010, gruber_direct_2018}, the flame front substantially modifies the incoming turbulent flow even on the reactants side, producing an abrupt shift in the characteristic turbulent time and length scales across the flame brush toward the products. In addition, the degree of strain imposed on the flame front varies with wall-normal distance, as anisotropic turbulent motions of different intensities interact with it. These combined effects lead to distinct flame characteristics depending on proximity to the walls.
The following discussion only focuses on the flame at $\varphi = 0.35$, which frequently propagates toward the channel boundaries, in contrast to the leaner flame at $\varphi = 0.25$ that remains predominantly confined to the core flow.


\subsubsection{Flame interaction with the turbulent boundary layer}

\begin{figure}[ht!]
    \centering
    \includegraphics[width=0.9\textwidth]{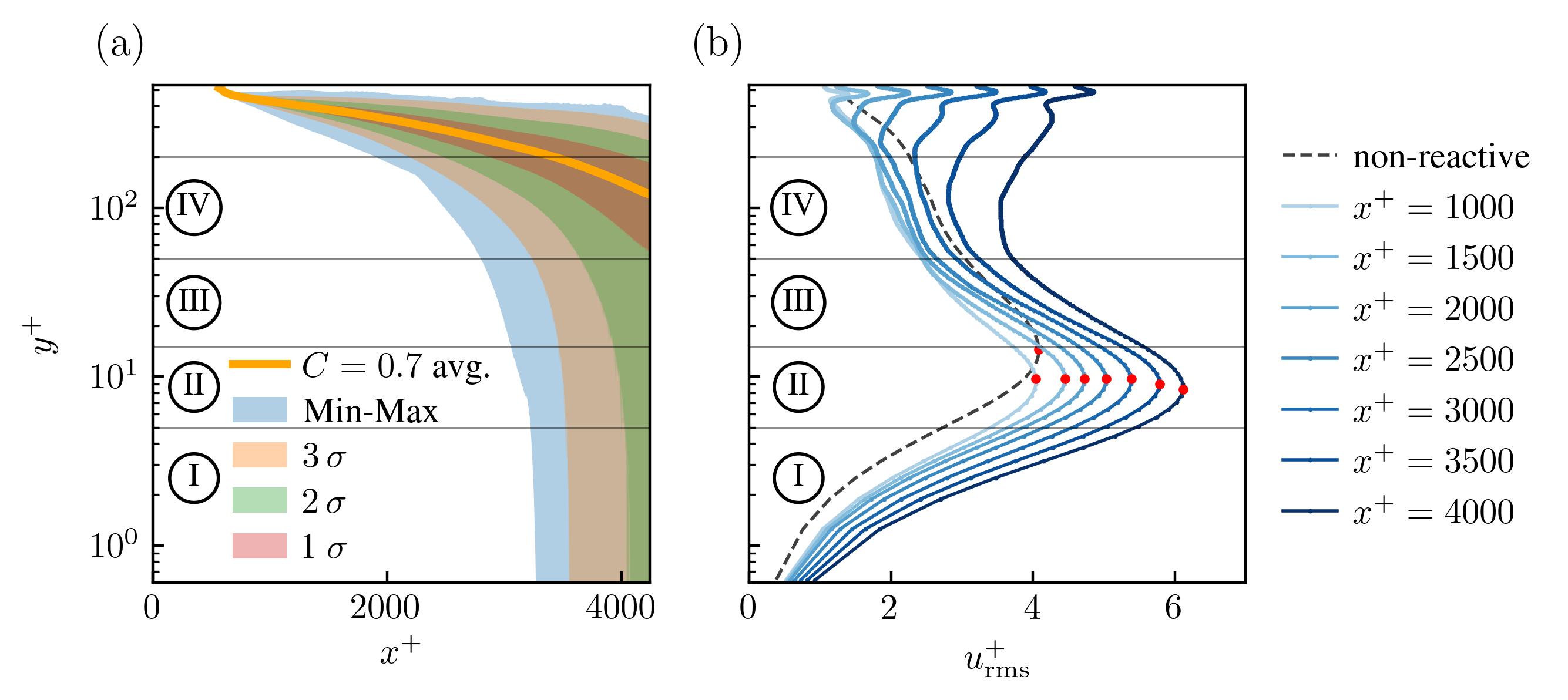}
    \caption{$(a)$ Statistics of the instantaneous flame position of the $\varphi=0.35$ flame in the physical space $y^+$ over $x^+$. $(b)$ Blue curves show the $u^+_\mathrm{rms}$ profiles on the abscissa as function of the wall distance $y^+$ on the ordinate for different streamwise locations $x^+$ within the reactive DNS. The maximum value of each profile is marked in red and the black dashed line denotes the profile of the inert precursor DNS.
    }
    \label{fig: flame boundary layer}
\end{figure}
To confirm that, in the $\varphi=0.35$ flame configuration, a sufficient number of flame samples are present within the near-wall region of the turbulent boundary layer, the statistics of the instantaneous flame-surface position (described by $C_\mathrm{norm}=0.7$) are shown in figure~\ref{fig: flame boundary layer}$(a)$. On average, the flame front resides at a wall distance of approximately $y^+ \approx 100$. However, its fluctuations—within two and three standard deviations of the mean—extend all the way to the solid walls for streamwise positions $x^+ \gtrsim 3500$. The overall minimum and maximum instantaneous flame locations are also indicated, enclosing the range corresponding to three standard deviations about the mean.
The strength of the turbulent motions interacting with the flame front is characterized in figure~\ref{fig: flame boundary layer}$(b)$, which presents the dimensionless rms velocity fluctuation ($u^+_\mathrm{rms}$) as a function of wall-normal distance $y^+$, colored by streamwise position $x^+$. Red markers highlight the peak of each $u^+_\mathrm{rms}$ profile. For comparison, the corresponding non-reactive precursor DNS is shown as a black dashed line, illustrating the compression of the turbulent boundary layer caused by the flame. This effect was previously reported by \citet{gruber_turbulent_2010, gruber_direct_2018}.
Guided by the location of the peak turbulent fluctuations, four spatially-segregated bins, labeled $\mathrm{I}$–$\mathrm{IV}$ in figure~\ref{fig: flame boundary layer}, are defined for further analysis in the wall-normal direction.


\begin{figure}[ht!]
    \centering
    \includegraphics[width=1.05\linewidth]{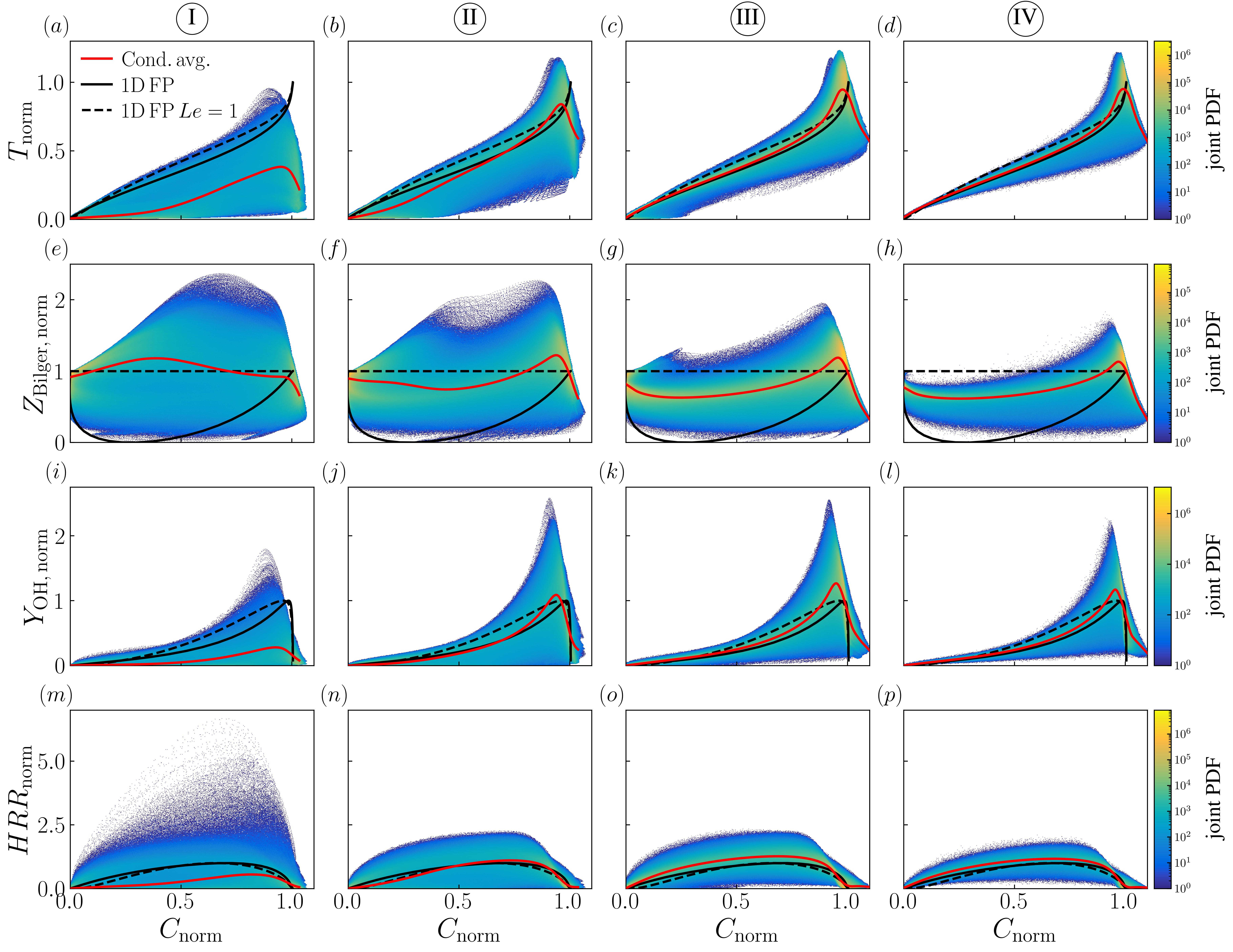}
    \caption{Joint PDF of the thermochemical state (normalized temperature $T_\mathrm{norm}$ in $(a$-$d)$, mixture fraction $Z_\mathrm{Bilger,\,norm}$ in $(e$-$h)$, \ce{OH}-radical mass fraction $Y_{\ce{OH}.\,\mathrm{norm}}$ in $(i$-$l)$ and heat release rate $\mathrm{HRR}_\mathrm{norm}$) in $(m$-$p)$ of the $\varphi=0.35$ flame over the normalized progress variable $C_\mathrm{norm}$ for the bins $\mathrm{I}$-$\mathrm{IV}$ in wall-normal direction, corresponding to figure~\ref{fig: flame boundary layer}.
    The red line denotes the conditional average, while black lines show the 1D laminar FP profiles with detailed transport (straight line) and unity Lewis number assumption (dashed line).
    }
    \label{fig: ybins flame structure}
\end{figure}
Variations in the local flame structure as a function of wall-normal distance are examined using bins $\mathrm{I}$–$\mathrm{IV}$ (see figure~\ref{fig: flame boundary layer}). Figure~\ref{fig: ybins flame structure} presents joint PDFs of the thermochemical state—normalized temperature $T_\mathrm{norm}$ $(a$–$d)$, normalized mixture fraction $Z_\mathrm{Bilger,\,norm}$ $(e$–$h)$, normalized \ce{OH} mass fraction $Y_{\ce{OH},\,\mathrm{norm}}$ $(i$–$l)$, and normalized heat-release rate $\mathrm{HRR}_\mathrm{norm}$ $(m$–$p)$—as functions of the normalized progress variable $C_\mathrm{norm}$. Conditional averages are plotted in red, while solutions of homologous 1D laminar FP flames calculated using mixture-averaged transport (including the Soret effect) and unity Lewis number assumptions appear as black reference lines.

Bin~$\mathrm{I}$, located closest to the wall, displays markedly different thermochemical behavior and will therefore be discussed separately. Bins~$\mathrm{IV}$–$\mathrm{II}$, on the other hand, quantitatively describe the monotonic transition from the core flow to the near-wall region. As the flame moves toward the wall—from bin~$\mathrm{IV}$ (farthest from the wall) to bin~$\mathrm{II}$—the turbulence intensity increases correspondingly. Across all bins, the combined influence of turbulence and TD phenomena manifests as broad state-space distributions: substantial mixture-fraction variability, super-adiabatic temperature excursions, elevated radical concentrations, and enhanced heat-release rates. With increasing turbulence levels for decreasing wall-normal distance, fluctuations about the conditional means also grow, consistent with expectations for stronger turbulent forcing.
Comparison of the conditional averages (red lines in figure~\ref{fig: ybins flame structure}) with the 1D laminar FP-flames solutions (black lines) shows the closest agreement in bin~$\mathrm{IV}$ (panels $(d,h,l,p)$), which corresponds to the core-flow region. The similarity between the turbulent and laminar thermochemical states progressively deteriorates when moving toward the wall through bins~$\mathrm{III}$–$\mathrm{I}$. This trend reflects the increasing turbulence intensity acting on the flame front in bins~$\mathrm{III}$ and $\mathrm{II}$, and the growing influence of heat loss in bin~$\mathrm{I}$.

Bin~$\mathrm{I}$ lies very close to the wall ($y^+ \leq 5$), within the viscous sublayer, where FWI significantly alters the flame structure. The conditional averages exhibit markedly reduced temperature and heat-release rate, together with a substantial spread in mixture fraction. Intermittent quenching of the flame on the solid surfaces also occurs in this region. A more detailed analysis of the observed FWI and intermittent flame quenching is provided in section~\ref{sec: results FWI}.

\subsubsection{Stretch factor $I_0$}

Because the flame front only intermittently reaches wall-normal distances below $y^+\approx100$, (figure~\ref{fig: flame boundary layer},(a)), the methodology of \citet{howarth_empirical_2022} and \citet{Day2009} is adapted to evaluate the stretch factor $I_0$ in bins $\mathrm{I}$-$\mathrm{IV}$.
In this approach, flame trajectories, interpretable as “flamelets” \citep{Peters1988}, are extracted using the direction of the progress-variable gradient. For each trajectory, the local thermal flame thickness,
$\delta_F^*=(T_\mathrm{b}-T_\mathrm{u})\big/\max(\mathrm{d}T/\mathrm{d}l)$
and the local consumption speed,
$s_c^*=-1/(\rho_\mathrm{u} Y_\mathrm{\ce{H2},\,u}) \int_l \dot{\omega}_\mathrm{\ce{H2}} \mathrm{d}l$
evaluated along the trajectory $l$, are determined.
\begin{figure}[ht!]
    \centering
    \includegraphics[width=0.85\linewidth]{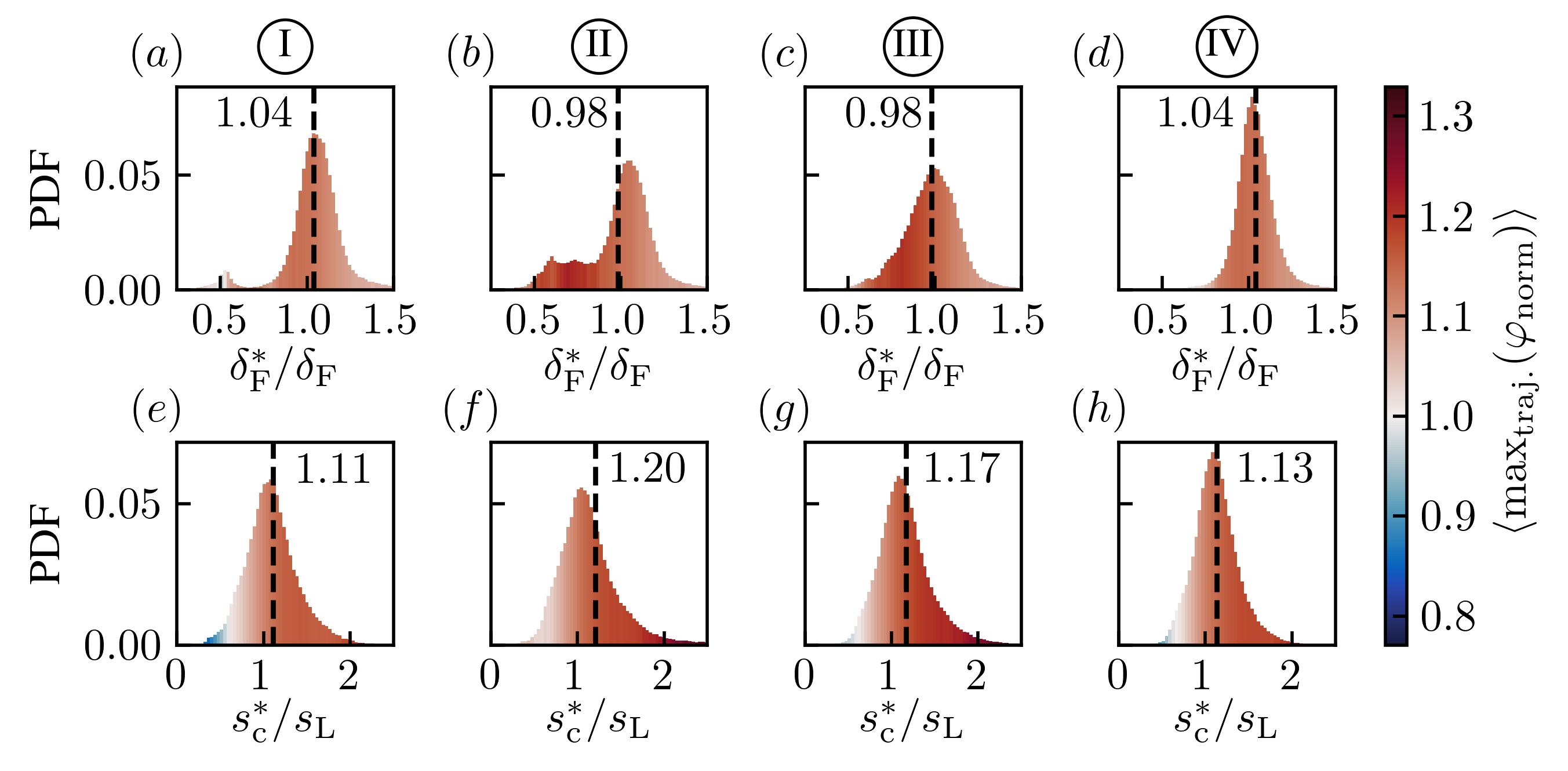}
    \caption{Distribution of the local flame thickness $(a$-$d)$ and local consumption speed $(e$-$h)$ for the bins $\mathrm{I}$-$\mathrm{IV}$ in wall-normal direction, for the $\varphi=0.35$ flame. The histograms are colored by the average trajectory-wise maximum equivalence ratio and the dashed line annotates the mean value of each distribution.}
    \label{fig: ybins delta_F sC}
\end{figure}
Figure~\ref{fig: ybins delta_F sC},(a–d) presents the histogram distributions of the normalized local flame thickness for bins $\mathrm{I}$-$\mathrm{IV}$. Although the mean flame thickness remains close to the laminar reference in all wall-normal regions, the distributions broaden and become increasingly bimodal when moving from the core flow (bin~$\mathrm{IV}$, panel (d)) toward the wall (bin~$\mathrm{I}$, panel (a)). The coloring, based on the normalized maximum equivalence ratio along each trajectory, shows that, in bins $\mathrm{II}$ and $\mathrm{III}$ (panels (b) and (c)), thinner flames tend to be associated with locally enriched mixtures.
The distributions of normalized local consumption speed are shown in figure~\ref{fig: ybins delta_F sC},(e–h). The ensemble-averaged value across the trajectories yields the stretch factor,
$I_0=\langle s_\mathrm{c}/s_\mathrm{L}\rangle_\mathrm{traj.}$.
As the flame approaches the wall, the distributions widen and extend to larger values, increasing from bin~$\mathrm{IV}$ (panel (h)) toward bin~$\mathrm{II}$ (panel (e)). The color coding further highlights the correlation between local equivalence ratio and consumption speed: enriched mixtures produce higher trajectory flame speeds, whereas leaner mixtures result in reduced values.
\begin{figure}[ht!]
    \centering
    \includegraphics[width=0.8\linewidth]{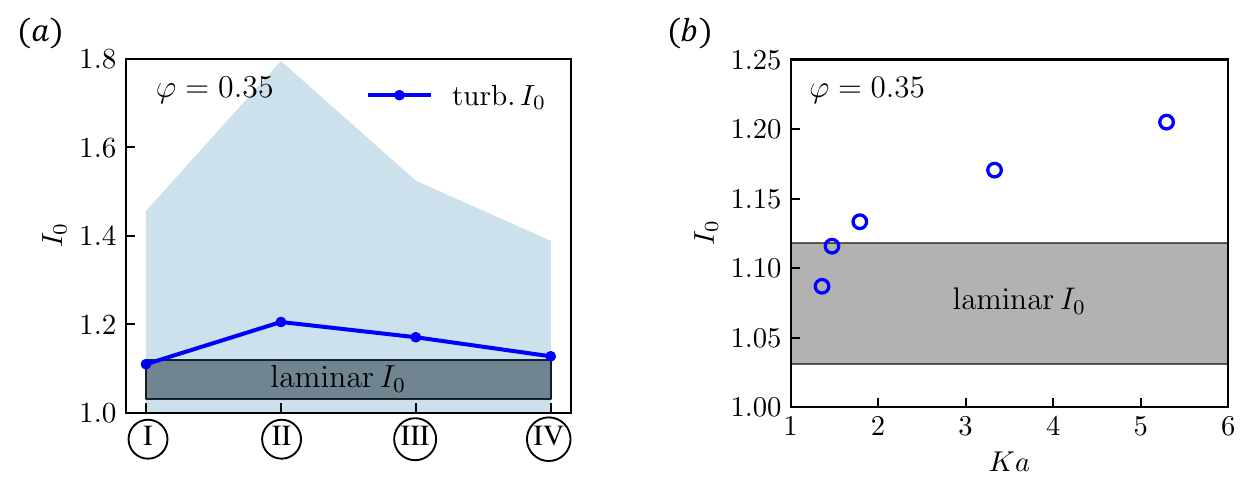}
    \caption{Averaged stretch factor $I_0$ in bins $\mathrm{I}$-$\mathrm{IV}$ corresponding to different wall-normal distances $(a)$. The blue corridor denotes one standard deviation. For reference, the range of $I_0$ values for homologous laminar flames is marked by the gray shaded area. Local stretch factor $I_0$ plotted against the local Karlovitz number $(b)$.}
    \label{fig: I0 ybins}
\end{figure}
Figure \ref{fig: I0 ybins}(a) shows the stretch factor $I_0$, together with one standard deviation, evaluated in bins~$\mathrm{I}$-$\mathrm{IV}$ as a function of wall-normal position. A clear increase in $I_0$ is observed when moving from the core flow toward the wall, with both the mean and fluctuation levels reaching a pronounced peak in bin $\mathrm{II}$, which aligns with the location of maximum turbulence intensity. For reference, the range of $I_0$ values corresponding to the homologous laminar flames is indicated by the gray-shaded region, highlighting the substantial synergistic interaction between turbulence and TD phenomena. These combined effects strengthen with increasing turbulence intensity, leading to elevated stretch-factor values particularly in bins $\mathrm{II}$ and $\mathrm{III}$.


As shown in figure~\ref{fig: flame boundary layer}$(a)$, the flame-front position exhibits strong wall-normal fluctuations, which substantially influence the flame’s response to the combined effects of turbulent forcing and TD phenomena. Consequently, the $I_0$ estimates reported in figure~\ref{fig: sC AT I0 xbins}$(d)$, computed from quantities sampled in longitudinal bins extending across the wall-normal direction, tend to be lower than those in figure~\ref{fig: I0 ybins}$(a)$, where the bins are defined based on wall distance alone. For instance, the peak value of the averaged stretch factor $I_0$ is approximately 1.16 in the former case, compared to about 1.2 in the latter. This difference provides additional evidence that the strongest synergistic interactions between TD effects and boundary-layer turbulence arise within the buffer layer, where they coincide with the peak intensities of shear-driven, anisotropic turbulent motions (bin $\mathrm{II}$).

Importantly, the present DNS analysis establishes a quantitative correlation between the local Karlovitz number, defined by the black dots in figure~\ref{fig: Karlovitz}, and the local stretch factor $I_0$, shown in figure~\ref{fig: I0 ybins}$(b)$ in a shear-driven, wall bounded turbulent channel flow flame. At relatively low Karlovitz numbers ($Ka \sim 1$–$2$), $I_0$ rises steeply, exceeding its laminar counterpart for $Ka \gtrsim 1.8$. At elevated Karlovitz numbers ($Ka \sim 2$–$6$), the increase in $I_0$ becomes more gradual and approaches a linear trend. 
In summary, the present results demonstrate a monotonic increase of the local stretch factor $I_0$ with increasing local Karlovitz number $Ka$. This behavior is consistent with earlier findings in other turbulent-flame configurations, such as the turbulent slot flames studied by \citet{berger_effects_2024} and the statistically planar flames interacting with forced isotropic turbulence examined by \citet{lu_modeling_2021}, \citet{yao_isolating_2024} and \citet{hunt_thermodiffusively-unstable_2025}.


\subsubsection{Flame--wall interaction} \label{sec: results FWI}

The channel walls, which confine the present turbulent reacting-flow configuration, influence not only the characteristics of the turbulent flow that approaches the flame from the unburned side, but also its propagation in their immediate vicinity. This dual effect is qualitatively visible in the instantaneous fields of figure~\ref{fig: avg_instant_flame} and quantitatively supported by the statistics in figure~\ref{fig: ybins flame structure}$(a,c,i,m)$. At the wall, strong heat losses and localized quenching dominate the flame structure and its propagation behavior. Although a detailed investigation of flame--wall interaction (FWI) is not the primary objective of this work, this section offers a brief examination of the FWI processes affecting the portions of the premixed flame front that intermittently approach the solid boundaries.

FWI has traditionally been studied in two idealized premixed-flame configurations: head-on quenching (HOQ), where a planar flame propagates normal to the wall and quenches due to heat losses or reactant depletion, and side-wall quenching (SWQ), where a flame propagates parallel to the wall and is continuously quenched within a near-wall quenching layer. When the mean flame-propagation direction is neither normal nor parallel to the wall, as in the present rod-stabilized, V-shaped turbulent flame developing in a fully developed channel flow, the interaction is commonly referred to as oblique wall quenching (OWQ) \citep{Ahmed2020, Wang2023, Wang2025}.

Recently, \citet{de_nardi_infinitely_2024} introduced a simplified infinitely-fast catalytic-wall model to represent heterogeneous surface reactions during FWI in hydrogen–air mixtures. The model treats the wall as an efficient sink for radicals by assuming instantaneous adsorption of all hydrogen radicals present in the fluid cells adjacent to the surface. This approach represents the opposite limiting case to the widely used inert-wall assumption employed in earlier FWI studies \citep{bruneaux_flame-wall_1996, Dabireau2003, gruber_turbulent_2010, gruber_direct_2012, gruber_direct_2018}. As such, its adoption helps resolve uncertainties related to transient spikes in wall heat-release rate that may arise from recombination of overpredicted radical concentrations in stoichiometric or fuel-rich flames. However, this transient heat-release augmentation is not observed in fuel-lean 1D laminar FWI simulations, either by \citet{de_nardi_infinitely_2024} or in our own computations. Therefore, the standard assumption of isothermal, chemically inert walls is considered adequate to accurately represent the FWI process in the present work. Consistently, the 3D DNS data also show no evidence of heat-release enhancement at the wall during turbulent FWI, as demonstrated in figure~\ref{fig: avg_instant_flame_zoom}$(h)$.

\begin{figure}[ht!]
    \centering
    \includegraphics[width=1.05\textwidth]{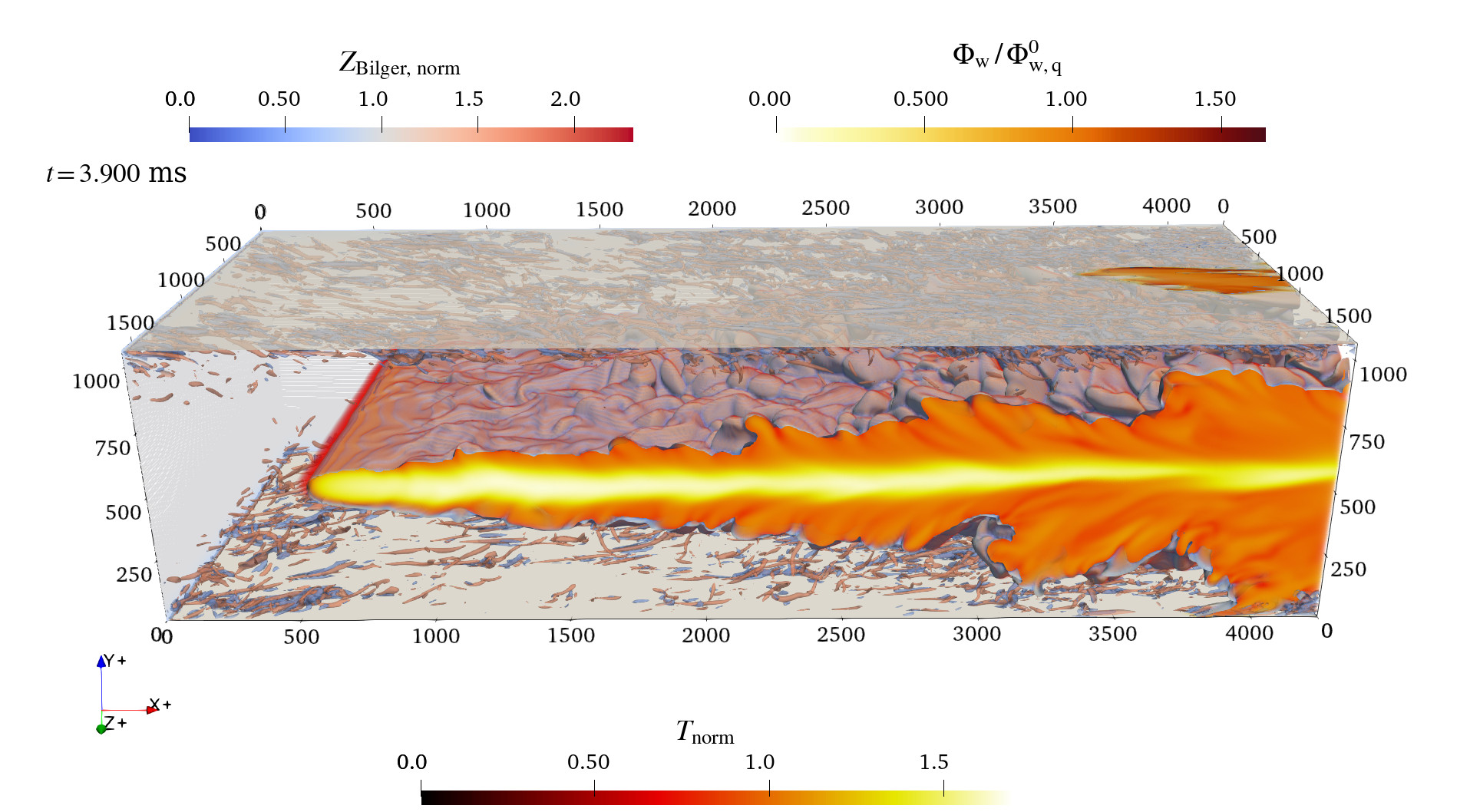}
    \vspace{1mm}
    \caption{
    Instantaneous snapshot of the $\varphi=0.35$ flame with the normalized temperature $T_\mathrm{norm}$ rendered as volume, normalized mixture fraction $Z_\mathrm{Bilger,\,norm}$ shown on the $C_\mathrm{norm}=0.7$ flame surface and the normalized WHF $\Phi_\mathrm{w}/\Phi_\mathrm{w,q}^0$ on the walls. Turbulent structures are illustrated by the $Q$-criterion, with red and blue iso-surfaces representing positive and negative iso-values, respectively.
    }
    \label{fig: 3d FWI flame}
\end{figure}
Figure \ref{fig: 3d FWI flame} shows an instantaneous snapshot of the V-shaped flame anchored in the TCF configuration at $\mathrm{Re}_\tau = 530$ and $\varphi = 0.35$. The 3D volume rendering displays the normalized temperature field, along with a flame-surface iso-contour at $C_\mathrm{norm}=0.7$, colored by the normalized Bilger mixture fraction $Z_{\mathrm{Bilger,\,norm}}$. Vortical structures are visualized using the $Q$-criterion, with red and blue iso-surfaces indicating positive and negative iso-values, respectively. The wall-heat flux (WHF) $\Phi_\mathrm{w}$, normalized by the quenching WHF of the corresponding 1D laminar-flame HOQ $\Phi_\mathrm{w,q}^0$, is represented by the color pattern on the solid walls. The WHF, evaluated at the wall (index $\mathrm{w}$), is defined as
\begin{equation}\label{eq: WHF}
    \Phi_\mathrm{w} = - \lambda \, \frac{\partial T }{ \partial x_i} \bigg|_\mathrm{w} \, n^\mathrm{w}_i \,\,,
\end{equation}
where $\lambda$ is the thermal conductivity and $n_i^\mathrm{w}$ the wall-normal unit vector. An animation of figure~\ref{fig: 3d FWI flame} is provided as Movie 2 in the supplementary material.

Under laminar conditions, SWQ and OWQ typically produce substantially lower WHF levels than HOQ \citep{heinrich_investigation_2018, Wang2023}. In contrast, WHF values observed in turbulent SWQ are comparable to those from 1D HOQ, making the latter a convenient normalization reference.
\begin{figure}[ht!]
    \centering
    \includegraphics[width=0.675\linewidth]{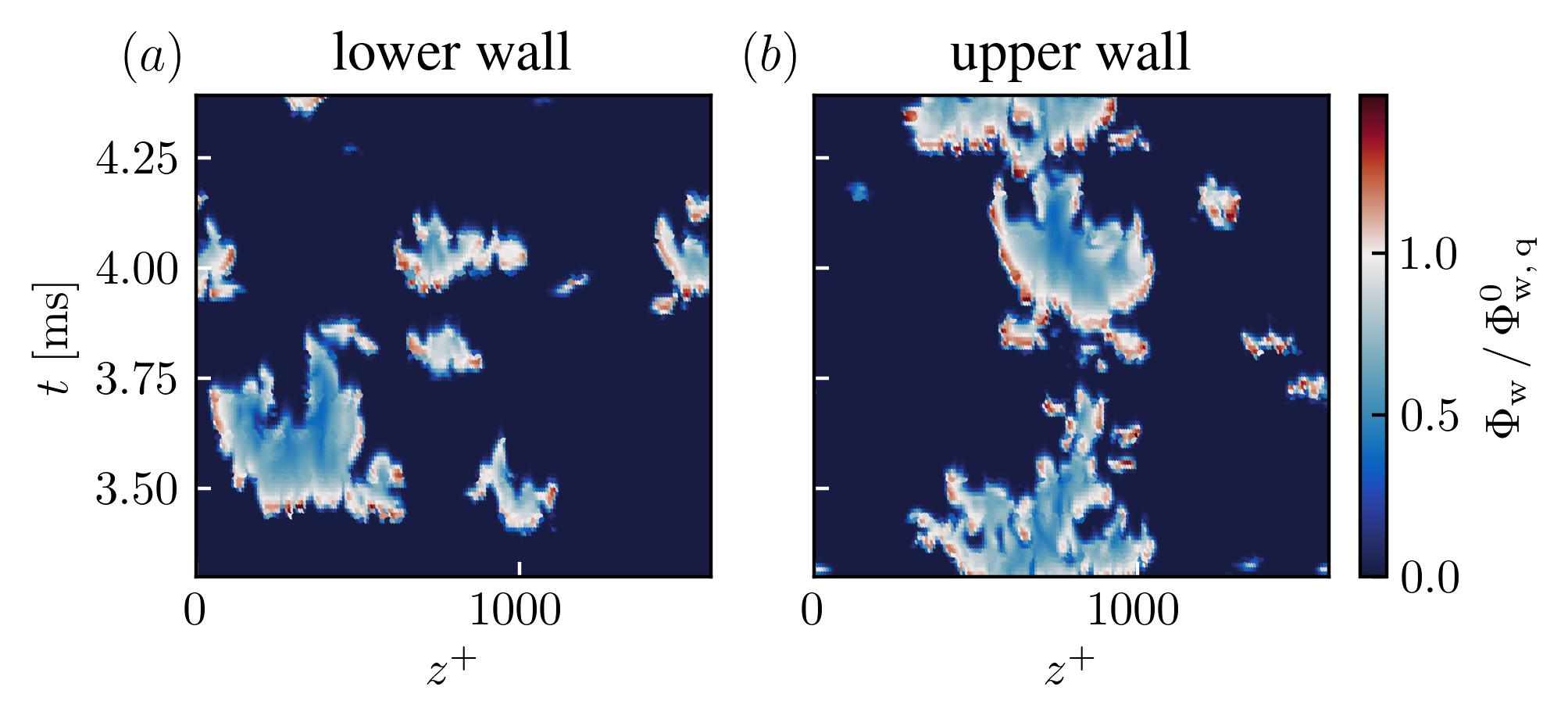}
    \caption{Spanwise and temporal values of the normalized WHF sampled at streamwise position $x^+=4200$ at the lower wall in $(a)$ and the upper wall $(b)$.}
    \label{fig: temporal spatial occurence WHF}
\end{figure}
Figure \ref{fig: temporal spatial occurence WHF} shows the temporal and spanwise ($z^+$) distribution of the normalized WHF at the streamwise location $x^+=4200$. These data confirm the occurrence of FWI and further demonstrate that local WHF values can exceed the 1D HOQ reference level. Additionally, Movie 3 in the supplementary material provides an animation of the instantaneous WHF at both walls, highlighting all instances of FWI.

For a more detailed characterization of the FWI process, the local flame-quenching point is extracted. Determining this position in turbulent FWI is non-trivial, as extrema of reaction rate and WHF are generally spatially and temporally de-coupled \citep{heinrich_investigation_2018, heinrich_large_2018}. In contrast to laminar flames—where several criteria based on peak WHF, reaction rate, consumption speed, or threshold values of temperature or radical concentration typically yield consistent results—turbulent flames require a more robust definition. Accordingly, an enthalpy-based quenching criterion is adopted: points on the $C_\mathrm{norm}=0.7$ flame surface are classified as quenched when their local enthalpy matches the quenching enthalpy of the 1D laminar HOQ reference flame. This approach provides a consistent identification of the quenching location, the corresponding quenching WHF ($\Phi_\mathrm{w,q}$), and the associated quenching distance ($x_\mathrm{q}$). Here, $\Phi_\mathrm{w}$ denotes the WHF defined in equation~\ref{eq: WHF}, whereas $\Phi_\mathrm{w,q}$ refers specifically to its value at the quenching point.

\begin{figure}[ht!]
    \centering
    \includegraphics[width=0.775\linewidth]{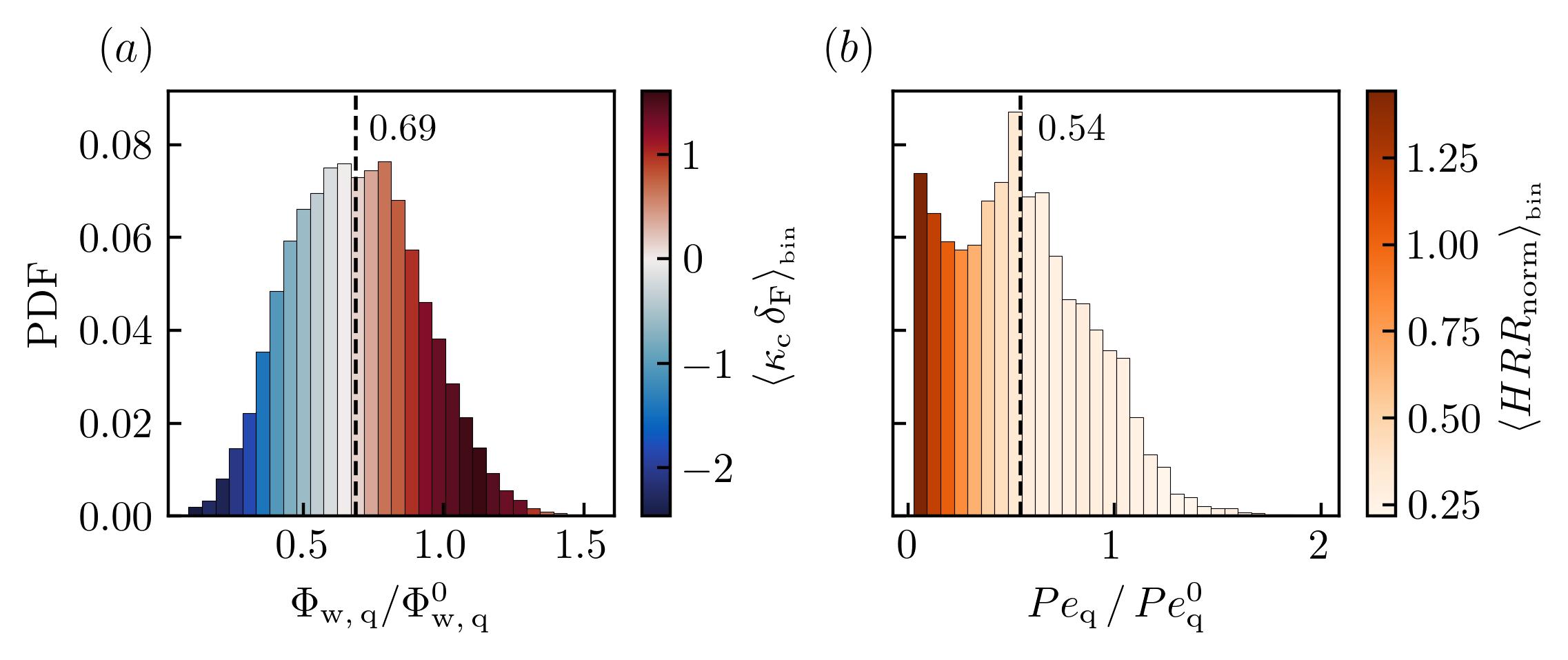}
    \caption{$(a)$ Histogram of the normalized quenching WHF $\Phi_\mathrm{w,\,q}/\Phi_\mathrm{w,\,q}^0$, colored by the bin-averaged normalized curvature $\kappa_\mathrm{c}\delta_\mathrm{F}$. $(b)$ Distribution of the normalized quenching Péclet-number ($Pe_\mathrm{q} = x_\mathrm{q} / \delta_\mathrm{F}$), colored by the bin-averaged normalized heat release rate $HRR_\mathrm{norm}$ of the quenching flame.}
    \label{fig: quenching WHF and xq}
\end{figure}
Figure~\ref{fig: quenching WHF and xq}$(a)$ shows histograms of the normalized quenching WHF, colored by the bin-averaged flame curvature. Owing to TD effects, positively curved regions of the flame front exhibit elevated WHF values, often exceeding the 1D laminar HOQ reference, because local fuel enrichment enhances the burning intensity. In contrast, negatively curved regions experience reduced WHF due to locally fuel-lean conditions. Figure~\ref{fig: quenching WHF and xq}$(b)$ presents the distribution of the normalized quenching Péclet number (the non-dimensional quenching distance, $Pe_\mathrm{q}=x_\mathrm{q}/\delta_\mathrm{F}$), colored by the bin-averaged normalized heat-release rate. The results clearly demonstrate that turbulent motions substantially reduce the quenching distance relative to the 1D laminar HOQ case, and that quenching occurring closer to the wall is associated with increased local and mean heat-release rates. This trend is likely linked to TD-induced modifications of the microscopic flame structure via mixture-fraction variance: localized fuel enrichment in positively curved regions enhances reactivity and flame power, enabling the flame to propagate closer to the wall at specific locations.

In summary, the present 3D DNS analysis shows good qualitative agreement with key findings from 2D laminar HOQ studies of flames affected by TD phenomena \citep{PartI, PartII}. TD effects manifest clearly in the turbulent FWI process, producing a broad range of WHF magnitudes: positively curved flame segments lead to locally enriched mixtures and elevated WHF, whereas negatively curved regions yield lower WHF due to localized fuel-lean conditions. Moreover, the quenching distance in the turbulent configuration is markedly smaller than in the laminar HOQ cases. This reduction has important practical implications, as it indicates potentially higher near-wall thermal loads and thus increased relevance for combustor-wall design and safety considerations.


\section{Conclusion and Outlook} \label{sec: conclusion}

The present study reports direct numerical simulations (DNS) of fuel-lean premixed hydrogen flames—susceptible to thermo-diffusive (TD) phenomena—anchored in fully developed turbulent channel flow (TCF) at a friction Reynolds number of $\mathrm{Re}_\tau = 530$. Two flame configurations are investigated: an ultralean premixed flame at an equivalence ratio of $\varphi = 0.25$, which predominantly propagates within the channel core, and a mildly stratified flame at $\varphi = 0.35$, which burns slightly faster and therefore advances closer to the walls, intermittently undergoing FWI and quenching. Both turbulent-flame configurations exhibit characteristic TD-driven features, including super-adiabatic temperatures, cellular flame structures, localized fuel enrichment and depletion, and pronounced spatial variations in local reactivity. With respect to the three research questions introduced earlier, the present analysis provides the following insights into the propagation of premixed hydrogen flames influenced by TD phenomena and interacting with realistic shear turbulence in a canonical fully developed TCF environment:

(i) Global flame-propagation characteristics. Both flames propagate at varying wall-normal distances and therefore experience different turbulence intensities and local Karlovitz numbers. These variations enhance flame stretch, increase flame-surface area, and ultimately raise the turbulent flame speed. The stretch factor $I_0$ increases in the streamwise direction and quantifies the flame-speed augmentation caused by locally enhanced reactivity induced by TD phenomena. This behavior indicates a synergistic interaction between turbulence and TD effects, with $I_0$ exceeding its laminar reference value. Notably, the ultra-lean $\varphi=0.25$ flame exhibits a stronger increase in $I_0$ and thus more pronounced synergistic effects, despite burning predominantly in the core region of the channel. Although TD instabilities are only weakly developed under the present conditions, characterized by the relatively high reactant preheat temperature typical of gas-turbine applications, the resulting increase in $I_0$ within the turbulent flames remains significant and carries important implications for modeling and practical deployment.

(ii) Flame topology and microstructure. Both flames exhibit comparable topological features, consisting predominantly of cylindrically curved or nearly planar flame elements, and display similar mean curvature radii when normalized by the thermal flame thickness. The most probable curvature radius is of the same order as the most unstable wavelength, the critical wavelength, and the Taylor microscale. Moreover, the microscopic flame structure shows a clear correlation with the local flame curvature, highlighting the central role of curvature in turbulent-combustion modeling. Downstream evolution reveals similar trends in the development of TD instabilities for both flames, with the leaner case ($\varphi = 0.25$) exhibiting more pronounced manifestations, including higher normalized super-adiabatic temperatures, enhanced reactivity, and stronger mixture enrichment.

(iii) Effects of the wall-normal distance on flame propagation. As the $\varphi = 0.35$ flame propagates from the core flow toward the walls, it experiences increasing turbulence intensities and a broadening range of local Karlovitz numbers, which amplify fluctuations in the thermochemical state. The stretch factor $I_0$ rises systematically with the local Karlovitz number and reaches its maximum within the buffer layer, where the most intense unsteady motions and velocity fluctuations occur. The fact that $I_0$ exceeds its laminar reference value provides clear evidence of synergistic interactions between turbulent boundary-layer dynamics and TD phenomena. This behavior is particularly significant for practical applications, as it relates directly to the risk of boundary-layer flashback in hydrogen-fired combustors.



Further work should investigate turbulent propagation of premixed hydrogen-air flames susceptible to stronger TD phenomena and shear-driven turbulent intensity, respectively targeting higher pressures and Reynolds numbers, to broaden the parameter space of flow and thermochemical conditions and improve the generality of the present conclusions. In particular, analyses based on different ratios between turbulence length scales and intrinsic flame instability wavelengths may provide additional insights into the mechanisms that control propagation of TD-susceptible premixed flames in turbulent boundary layers.

\subsection*{Supplementary data}
The following supplementary/ancillary material are provided:
\\
\textbf{Movie 1 \& 2}: 3D animations of the turbulent channel flow flames $\varphi=0.25$ and $\varphi=0.35$, respectively, with volume rendering of the normalized temperature $T_\mathrm{norm}$. The iso-surface $C_\mathrm{norm}=0.7$ is colored by the normalized mixture fraction $Z_\mathrm{Bilger,\,norm}$, the upper and lower walls of the channel flow show the normalized wall heat flux $\Phi_\mathrm{w}/\Phi_\mathrm{w,\,q}^0$, and turbulent eddies are illustrated by the $Q$-criterion.
\\
\textbf{Movie 3}: Temporal animation of the normalized wall heat flux $\Phi_\mathrm{w}/\Phi_\mathrm{w,\,q}^0$ at the upper and lower walls of the turbulent channel flow for the flame $\varphi=0.35$.
\\
\textbf{Supplementary Document}: Dispersion relation and stretch factor from the respective homologous two-dimensional laminar unstable reference flames. 

\subsection*{Acknowledgments}
The DNS calculations were performed in Norway on the BETZY High-Performance Computing (HPC) facility managed by UNINETT Sigma2 - the National Infrastructure for High Performance Computing and Data Storage. In Germany, the DNS dataset was analyzed on the Lichtenberg high-performance cluster at the Technical University of Darmstadt.
Felix Rong and Max Schneider thank Dr.-Ing. Matthias Steinhausen for his supervision and the fruitful discussions.

\subsection*{Funding}
The research work performed in Norway was initiated with financial support by the NCCS Centre, performed under the Norwegian research program Centres for Environment-friendly Energy Research of the Research Council of Norway (257579/E20). The computational and data storage allocations on the BETZY supercomputer were granted to the authors by UNINETT Sigma2 - the National Infrastructure for High Performance Computing and Data Storage in Norway (project numbers nn9527k and ns9121k). The analysis leading to these results has received funding from the European Research Council under the European Union’s Horizon Europe research and innovation program under the ERC synergy grant agreement no. 101119058 (HYROPE).
In Germany, this work has been funded by the Deutsche Forschungsgemeinschaft (DFG, German Research Foundation) – Project Number 523792378 - SPP 2419.

\subsection*{Declaration of interests}
The authors report no conflict of interest.

\subsection*{Author contributions}
\textbf{Felix Rong}: Conceptualization, Data Curation, Formal analysis, Methodology, Software, Visualization, Writing - Original Draft.
\textbf{Max Schneider}: Conceptualization, Methodology, Visualization, Writing - Review \& Editing.
\textbf{Hendrik Nicolai}: Conceptualization, Methodology, Project administration, Supervision, Writing - Review \& Editing.
\textbf{Christian Hasse}: Conceptualization, Funding acquisition, Resources, Supervision, Writing - Review \& Editing.
\textbf{Andrea Gruber}: Conceptualization, Data Curation, Funding acquisition, Investigation, Methodology, Project administration, Resources, Software, Supervision, Writing - Review \& Editing.

\bibliographystyle{jfm}
\bibliography{jfm}  






\end{document}